\documentclass[a4paper,11pt]{article}
\usepackage{hyperref}
\usepackage{xcolor}
\definecolor{terracotta}{RGB}{204,78,51}

\usepackage{latex_packages/jheppub,multirow}
\usepackage{mathtools}
\usepackage{physics}
\usepackage{float}
\usepackage{graphicx}
\usepackage{url}
\usepackage{cancel}
\usepackage{slashed}
\usepackage{placeins}
\usepackage{latex_packages/feynman}
\usepackage{soul} 
\usepackage{bm}
\usepackage{amsmath}
\usepackage{amssymb}
\usepackage{listings}
\usepackage[toc, page]{appendix}
\usepackage{color}
\usepackage{booktabs} 
\usepackage{pgf}
\usepackage{empheq}
\usepackage{tikz}
\usepackage[font=small,labelfont=bf,labelsep=period]{caption}
\usepackage{tabularx}
\newcolumntype{C}[1]{>{\centering\arraybackslash}p{#1}}
 \usetikzlibrary{decorations.pathmorphing}
  \usetikzlibrary{decorations.markings}
\tikzset{snake it/.style={decorate, decoration=snake}}
\usetikzlibrary{shapes.geometric, arrows}
\tikzstyle{process} = [rectangle, minimum width=3cm, minimum height=1cm, text centered, draw=black, fill=orange!30]
\tikzstyle{arrow} = [thick,->,>=stealth]
\usetikzlibrary{arrows.meta,bending}
\usepackage{subfigure}
\usepackage{pgf}
\usetikzlibrary{arrows}
\usetikzlibrary{patterns}
 \usepackage{braket}
 \usepackage[inline]{enumitem} 
 
\usepackage{pifont}
\usepackage{array}
\newcolumntype{P}[1]{>{\centering\arraybackslash}p{#1}}
\usepackage{amsthm}
\usepackage{hyperref}
 \hypersetup{
    colorlinks=true,
    linkcolor=blue,
    filecolor=black,      
    urlcolor=blue,
}
\usepackage{tensor}
\usepackage{slashed} 

\usepackage[dvipsnames]{xcolor}

\definecolor{violet}{cmyk}{0,1,0,0.2}

\newcommand{\be}{\begin{equation}}
\newcommand{\ee}{\end{equation}}
\newcommand{\bea}{\begin{eqnarray}}
\newcommand{\eea}{\end{eqnarray}}
\newcommand{\bi}{\begin{itemize}}
\newcommand{\ei}{\end{itemize}}
\newcommand{\ben}{\begin{enumerate}}
\newcommand{\een}{\end{enumerate}}

\def\K*{K^{(\ast)}}

\def\BtoK{B\rightarrow K}
\def\BtoK*{B\rightarrow K^{(\ast)}}

\numberwithin{equation}{section}
\numberwithin{figure}{section}
\numberwithin{table}{section}

\title{Descoped and Upscoped FCC-ee Running Scenarios in the SMEFT}
\author[a]{Eugenia Celada}
\author[b]{, Elie Hammou}
\author[c]{, Jaco ter Hoeve}
\author[d]{and Marion O.A. Thomas}

\preprint{MPP-2026-129}

\affiliation[a]{Department of Physics, University of Cyprus, Panepistimiou Street 1, Aglantzia, CY-1678 Nicosia, Cyprus}
\affiliation[b]{Nikhef Theory Group, Science Park 105, 1098 XG Amsterdam, The Netherlands}
\affiliation[c]{The Higgs Centre for Theoretical Physics, University of Edinburgh,\\[0.1cm]
JCMB, KB, Mayfield Rd, Edinburgh EH9 3FD, Scotland}
\affiliation[d]{Max-Planck-Institut für Physik, Boltzmannstrasse 8, 85748 Garching, Germany}

\emailAdd{celada.eugenia@ucy.ac.cy}
\emailAdd{ehammou@nikhef.nl}
\emailAdd{jaco.ter.hoeve@ed.ac.uk}
\emailAdd{mthomas@mpp.mpg.de}

\abstract{We present projections for the sensitivity to new physics of descoped and
upscoped FCC-ee scenarios in the framework of the Standard Model
Effective Field Theory (SMEFT). Starting from the baseline FCC-ee programme,
we analyse the subsequent impact of, first, removing the top-quark run, then
reducing the beam power from 50 MW to 30 MW, and finally removing two
interaction points (IPs), in all cases propagating the resulting loss in
top-quark mass precision through increased parametric uncertainties. We also
analyse the possibility of a staged top-quark run, where the top-quark run is partially
restored in case funding becomes available only at a later stage. Motivated by
the trade-off between beam power and operational costs, we also consider two
upscoped scenarios, either through a uniform increase in luminosity across all energy runs, or through a luminosity enhancement targeting exclusively the top-quark run. We explore the sensitivity and limitations
of each scenario and highlight in particular the complementarity between the FCC-ee projections and the (HL)-LHC measurements. 
By comparing the physics impact of the different running scenarios with the FCC-ee baseline, we show explicitly the importance of the top-quark run, indicating that a substantial fraction of the FCC-ee physics potential depends on it.
}
\begin{document}

\maketitle
\flushbottom

\section{Introduction}
\label{sec:intro}
The 2026 Update on the European Strategy for Particle Physics (ESPPU26) was adopted by the CERN Council in May 2026 and defines the strategic priorities for the future of European particle physics~\cite{CERN-ESU-2025-002}. In particular, the electron-positron Future Circular Collider (FCC-ee)~\cite{FCC:2025lpp,FCC:2025uan} has been identified as the preferred option for the next flagship collider to be hosted at CERN. The FCC-ee, through an intensive programme of runs at various centre-of-mass energies, would shed light on the properties of the Higgs, electroweak, and top sectors, and would pave the way towards a hadron-hadron collider, the FCC-hh. The ESPPU26 further identified a descoped FCC-ee as the preferred alternative option, should the FCC-ee prove financially unfeasible.

A descoped FCC-ee would be characterised by the construction of two interaction
points (IPs) instead of four, the removal of the run at the top-pair threshold,
and a reduced synchrotron radiation (SR) power \cite{Fabiola2025, Blondel:2026mdc, deBlas:2026:ESG2b}. These measures would lower the
construction cost by around 15\% \cite{Fabiola2025}, but the baseline physics programme of the
FCC-ee would be affected, as the instantaneous luminosity would be reduced.
One option to recover statistics is to run the descoped FCC-ee for a longer
period. The baseline FCC-ee is scheduled to run for four years at and around the
$Z$ pole, two years at the $WW$ threshold, three years at the $Zh$ threshold,
and five years at and above the top-pair threshold. The first three stages
therefore amount to nine years of running below the top-pair threshold. For the
same nine-year running period, the descoped FCC-ee would collect only a fraction
$0.365 \pm 0.010$ of the baseline integrated luminosity \cite{Blondel:2026mdc}. The baseline
statistics below the top-pair threshold could nevertheless be recovered by
extending the programme to 23 years~\cite{Blondel:2026mdc}. However, this
luminosity-based comparison does not capture all relevant considerations.

First, the absence of a top-pair threshold scan would prevent a precise determination of the top-quark mass. Such a determination is essential to make the most of the $Z$-pole run, since without it the electroweak precision observables (EWPOs) would remain limited by the parametric uncertainty on the top-quark mass \cite{Defranchis:2025auz}. In addition, the top-quark run at $\sqrt{s} = 365$ GeV provides sensitivity to the Higgs self-coupling through the energy dependence of the $Zh$ cross section \cite{terHoeve:2025omu, Maura:2025rcv}. It also leads to precise measurements of the top-quark electroweak couplings which are needed to control the theoretical uncertainty on the top-quark Yukawa coupling determination at FCC-hh \cite{FCC:2025lpp}. A top-quark run should therefore be retained, at least as a staged component of the FCC-ee programme, to fully exploit the FCC measurements and maximise the overall physics reach of the programme.

Second, although a descoped FCC-ee would have a reduced construction cost, a longer running plan necessarily leads to higher operational costs, and preliminary estimates indicate that the total cost of a 23-year descoped FCC-ee, including construction and operation, could be larger than that of the baseline FCC-ee~\cite{ECFA118}. Conversely, increasing the SR power beyond the planned 50 MW for the FCC-ee and 30 MW for the descoped FCC-ee would allow the target integrated luminosity to be reached in a shorter running time, thus reducing the operational cost for a fixed physics outcome. Considering an increased SR power naturally poses the question of an ``upscoped'' FCC-ee: should the necessary funding be available and the associated technical challenges be addressed, a FCC-ee with a larger SR power could achieve the target integrated luminosity faster than the baseline FCC-ee. If operated for a longer period, such a machine could therefore collect a substantially larger integrated luminosity, providing another relevant benchmark scenario for assessing the physics impact of FCC-ee scoping choices.

Given the wide range of possible FCC-ee running scenarios, it is necessary to compare their physics performance within a consistent framework. In this work, we systematically study the physics impact of several FCC-ee
benchmark scoping scenarios in the Standard Model Effective Field Theory (SMEFT)~\cite{Brivio:2017vri,Isidori:2023pyp}. The SMEFT provides a model-independent description of possible heavy new
physics, whose characteristic scale lies above the energies directly probed by
the collider, through higher-dimensional operators built from the SM fields and
symmetries. Within this framework, new physics effects can be indirectly tested
through precision measurements at future colliders and eventually interpreted in
terms of UV extensions of the SM \cite{terHoeve:2023pvs, Armadillo:2026mvp,Allwicher:2024sso, Maura:2024zxz, Gargalionis:2024jaw}. The SMEFT therefore constitutes an ideal
framework for comparing potential collider scenarios and provides a useful
benchmark for their new physics reach.
Our approach uses the open-source SMEFiT framework~\cite{Hartland:2019bjb,vanBeek:2019evb,Ethier:2021bye,Ethier:2021ydt,Giani:2023gfq,Celada:2024mcf,terHoeve:2023pvs,terHoeve:2025gey,terHoeve:2025omu} to assess the impact of various FCC-ee benchmark scoping scenarios on global analyses of the SMEFT parameter space. Building on the recent SMEFiT analysis of future electron-positron facilities~\cite{Armadillo:2026mvp}, we implement projections for the FCC-ee scoping scenarios, and compare their sensitivity to new physics with those of the baseline FCC-ee and LEP3.

The structure of this paper is as follows. First, Sect.~\ref{sec:scenarios} sets
out the alternative descoped and upscoped FCC-ee running scenarios considered in this work. Sect.~\ref{sec:analysis} then
describes the setup of our analysis, including the treatment of experimental and
theoretical uncertainties. Sect.~\ref{sec:results} presents the main results,
where we compare the various running scenarios to the baseline FCC-ee scenario and to LEP3. Finally, we summarise our findings and conclude in Sect.~\ref{sec:conclusion}. We collect supplementary results in App.~\ref{app:detailed_results}.

\section{Alternative FCC-ee scenarios}
\label{sec:scenarios}
In this section we detail the different scoping scenarios that we consider. These scenarios are further summarised in Tables~\ref{tab:descoped_runs} and~\ref{tab:upscoped_runs}, which respectively present the descoped scenarios, where the baseline FCC-ee programme is reduced, and the upscoped scenarios, where improved assumptions are considered. Tables~\ref{tab:descoped_runs} and~\ref{tab:upscoped_runs}  additionally include the running scenarios of the baseline FCC-ee programme and of LEP3, which we use for comparison. We combine the information associated with the data taken at the top-threshold energy scans and above and denote this combination as ``$\sqrt{s}=365$~GeV'' in the following.

\begin{table}
\centering
\footnotesize
\setlength{\tabcolsep}{10pt} 
\renewcommand{\arraystretch}{1.5} 
\begin{tabularx}{\linewidth}{|X|c|c|c|c|}
\hline
Collider & \# IPs & Energy $(\sqrt{s})$ & Luminosity ($\mathcal{L}_{\text{int}}$) & \% of Baseline \\
\hline
\hline
\multirow{4}{*}{Baseline FCC-ee} & \multirow{4}{*}{4 IPs} & 91.2 GeV & 205 $\text{ab}^{-1}$ & \multirow{4}{*}{100\%} \\
 & & 161 GeV & 19.2 $\text{ab}^{-1}$ &  \\
 & & 240 GeV & 10.8 $\text{ab}^{-1}$ &  \\
  & &  365 GeV (*) & 3.12 $\text{ab}^{-1}$ &  \\
\hline
\multirow{4}{*}{2 IP descoped FCC-ee} & \multirow{4}{*}{2 IPs} & 91.2 GeV & 74.8 $\text{ab}^{-1}$ & \multirow{3}{*}{36.5\%} \\
 & & 161 GeV & 7.0 $\text{ab}^{-1}$ &  \\
 & & 240 GeV & 3.9 $\text{ab}^{-1}$ &  \\
 & & (365 GeV) & (1.14 $\text{ab}^{-1}$) & (36.5\%) \\
\hline
\multirow{4}{*}{4 IP descoped FCC-ee} & \multirow{4}{*}{4 IPs} & 91.2 GeV & 123 $\text{ab}^{-1}$ & \multirow{3}{*}{60\%} \\
 & & 161 GeV & 11.5 $\text{ab}^{-1}$ & \\
 & & 240 GeV & 6.5 $\text{ab}^{-1}$ & \\
 & & (365 GeV) & (1.9 $\text{ab}^{-1}$) & (60\%) \\
\hline
\multirow{3}{*}{LEP3} & \multirow{3}{*}{2 IPs} & 91.2 GeV & 48 $\text{ab}^{-1}$ & 23.4\% \\
 & & 160 GeV & 5.6 $\text{ab}^{-1}$ & 29.2\% \\
 & & 230 GeV & 2.304 $\text{ab}^{-1}$ & 21.3\% \\
\hline
\end{tabularx}
\vspace{0.2cm}
\caption{Overview of the descoped FCC-ee running scenarios considered in this work. From left to right, we indicate the running scenario, the number of interaction points (IPs), the center of mass energy $\sqrt{s}$, the integrated luminosity $\mathcal{L}_{\rm int}$, and its percentage relative to the Baseline FCC-ee scenario at the same energy point. The numbers in parentheses represent the possible top-quark run. (*) We also consider the Baseline FCC-ee scenario without top-quark run in order to quantify its impact.}
\label{tab:descoped_runs}
\end{table}

\subsection{A descoped FCC-ee}
\label{subsec:descoped_scenarios}
The descoped FCC-ee project, as presented in Ref.~\cite{Fabiola2025}, is characterised by the removal of the run at $\sqrt{s}=365$ GeV, the construction of two IPs instead of four, and the reduction of the SR power from 50 to 30 MW per beam. Removing two IPs leads to a reduction in instantaneous luminosity by a factor of 0.61, while the decreased beam power reduces the luminosity by a factor of 0.60 due to the linear scaling between luminosity and the SR power~\cite{Blondel:2026mdc}. Adjusting the overall running time of the descoped FCC-ee as well as a possible staged top-quark run then leads to a variety of possible benchmark scenarios, which we detail here and summarise in Table \ref{tab:descoped_runs}.

\begin{itemize}
\item \textbf{2-IP descoped FCC-ee.}
Assuming the same run time as the baseline FCC-ee, this scenario considers a reduced SR power of 30 MW, as well as two IPs instead of four such that its total instantaneous luminosity is reduced to 36.5\% of the baseline FCC-ee value.
The absence of the top-quark run, besides the resulting loss of data, prevents the determination of the top-quark mass, thus increasing the parametric uncertainties associated with the lower energy runs \cite{Defranchis:2025auz}.
\item \textbf{Long descoped FCC-ee.} It was shown in Ref.~\cite{Blondel:2026mdc} that running the descoped FCC-ee for 23 years instead of 9 would yield the same integrated luminosity as the baseline FCC-ee. In this scenario, the only remaining differences with respect to the baseline are the absence of observables at $\sqrt{s}=365$ GeV and the lack of determination of the top-quark mass, which enters the parametric uncertainties. In terms of physics reach, this scenario is equivalent to the FCC-ee without a top-quark run. In the following sections, the results of the long descoped FCC-ee are therefore not shown separately, as they coincide with those of the FCC-ee scenario without a top-quark run.
\item \textbf{4-IP descoped FCC-ee.} While the SR power and the duration of the descoped FCC-ee programme could be increased should funding become available, decreasing the number of IPs is more problematic, as the decision to build only two IPs is irreversible. It has been argued in Ref.~\cite{ESGRecommendationSlides} that four caverns are needed for the FCC-hh programme, and excavating them only after the electron-positron stage might lead to a higher overall cost, defeating the purpose of a descoped FCC-ee. For this reason we also study here a benchmark scenario with four IPs, no top-quark run, and a reduced SR power of 30 MW. We assume a 9-year running schedule, corresponding to the baseline FCC-ee programme without the top-quark run. This scenario therefore leads to a total luminosity of 0.6 times that of the baseline FCC-ee.
\end{itemize}
The aforementioned descoped scenarios could be supplemented by a staged top-quark run, should funding become available, which would enable a precise determination of the top-quark mass, thus reducing the parametric uncertainties entering the electroweak precision observables. We therefore consider, for the 2- and 4-IP descoped scenarios,
the impact of adding a staged top-quark run lasting 5 years, as in the baseline FCC-ee
programme. This corresponds to a top-quark run integrated luminosity of 0.365 times
the baseline value for the 2-IP scenario, and 0.6 times the baseline value for
the 4-IP scenario, as indicated in parentheses in Table \ref{tab:descoped_runs}.

\begin{table}
\centering
\footnotesize
\setlength{\tabcolsep}{10pt} 
\renewcommand{\arraystretch}{1.5} 
\begin{tabularx}{\linewidth}{|X|c|c|c|c|}
\hline
Collider & \# IPs & Energy $(\sqrt{s})$ & Luminosity ($\mathcal{L}_{\text{int}}$) & \% of Baseline \\
\hline
\hline
\multirow{4}{*}{Baseline FCC-ee} & \multirow{4}{*}{4 IPs} & 91.2 GeV & 205 $\text{ab}^{-1}$ & \multirow{4}{*}{100\%} \\
 & & 161 GeV & 19.2 $\text{ab}^{-1}$ &  \\
 & & 240 GeV & 10.8 $\text{ab}^{-1}$ &  \\
 & & 365 GeV & 3.12 $\text{ab}^{-1}$ &  \\
\hline
\multirow{4}{*}{Upscoped FCC-ee} & \multirow{4}{*}{4 IPs} & 91.2 GeV & 246 $\text{ab}^{-1}$ & \multirow{4}{*}{120\%} \\
 & & 161 GeV & 23.04 $\text{ab}^{-1}$ &  \\
 & & 240 GeV & 12.96 $\text{ab}^{-1}$ &  \\
 & & 365 GeV & 3.744 $\text{ab}^{-1}$ &  \\
\hline
\multirow{4}{*}{Upscoped FCC-ee, top only} & \multirow{4}{*}{4 IPs} & 91.2 GeV & 205 $\text{ab}^{-1}$ & \multirow{3}{*}{100\%} \\
 & & 161 GeV & 19.2 $\text{ab}^{-1}$ &  \\
 & & 240 GeV & 10.8 $\text{ab}^{-1}$ &  \\
 & & 365 GeV & 4.68 $\text{ab}^{-1}$ & 150\% \\
\hline
\end{tabularx}
\vspace{0.2cm}
\caption{Overview of the upscoped FCC-ee running scenarios considered in this work. From left to right, we indicate the running scenario, the number of interaction points (IPs), the center of mass energy $\sqrt{s}$, the integrated luminosity $\mathcal{L}_{\rm int}$, and its percentage relative to the baseline FCC-ee scenario at the same energy point.}
\label{tab:upscoped_runs}
\end{table}

\subsection{The upscoped option}
Here we present alternative running scenarios with an increased integrated luminosity relative to the baseline FCC-ee. These scenarios are motivated by the possibility that, if technologically feasible, operating at higher SR power would allow the baseline luminosity targets to be reached within a shorter run time. This would reduce operational costs despite higher construction costs \cite{ECFA118}, so that the resulting savings could be reinvested in an extended running programme, leading to a higher overall integrated luminosity. 

In this work, we consider two viable alternative scenarios, summarised in Table~\ref{tab:upscoped_runs}. First, an overall increase in integrated luminosity of $20\%$ across all energy runs, and second, a targeted increase in integrated luminosity of $50\%$ in the top-quark run only, while keeping the integrated luminosities of the lower-energy runs unchanged from those in the original FCC-ee baseline scenario. This second scenario is motivated by the important role of the top-quark run in the FCC-ee physics programme and is designed to assess whether additional statistics at $\sqrt{s} = 365$ GeV would lead to a significant improvement in physics reach, in particular due to a more precise determination of the top-quark mass and of the top-quark electroweak couplings, and an improved sensitivity to the Higgs self-interactions. Comparing both upscoped scenarios allows us to study whether a targeted increase in the statistics of the highest-energy run is more beneficial than a uniform luminosity increase across the full FCC-ee programme.

\section{Analysis setup}
\label{sec:analysis}

In this section we summarise the theoretical and experimental settings of our analysis.
These settings closely follow recent SMEFiT analyses~\cite{Celada:2024mcf, Armadillo:2026mvp}. For completeness we summarise here the key points and refer the reader to the aforementioned studies for more details.

\paragraph{Theoretical framework}
\label{subsec:theoryframework}

 We consider CP-even dimension-six SMEFT operators in the Warsaw basis~\cite{Grzadkowski:2010es} and enforce the following flavour symmetry:
\begin{equation}
    U(2)_{q_L} \times U(2)_{u_R} \times U(3)_{d_R} \times U(3)_\ell \times U(3)_e \, .
\end{equation}
In total we consider $n_\text{op}=61$ independent operators, see App.~A in Ref.~\cite{Armadillo:2026mvp} for their definitions. 
We keep explicit the factors of the EFT cut-off scale $\Lambda$ in the definition of the Wilson coefficients, such that the dimension-six SMEFT Lagrangian can be written as:
\begin{equation}
    \mathcal{L}_{\text{SMEFT}} = \mathcal{L}_{\text{SM}} + \sum^{n_\text{op}}_{i=1} \frac{c_i}{\Lambda^2} \mathcal{O}_i^{(6)},
\end{equation}
where $\mathcal{O}_i^{(6)}$ are the dimension-six operators with associated dimensionless Wilson coefficients $c_i$. We adopt the $\{M_W, M_Z, G_F \}$ electroweak input scheme.
Furthermore, one-loop RGE effects in the strong and electroweak couplings are included for all operators. We adopt an initial scale of $\mu_0=10$ TeV and run down the coefficients to the scale of each observable \cite{terHoeve:2025gey}.

\paragraph{Experimental projections}
\label{subject_exp_projection}
The input datasets used for the (HL)-LHC, FCC-ee, and LEP3 follow those adopted in Ref.~\cite{Armadillo:2026mvp}.
Statistical uncertainties for the descoped and upscoped FCC-ee scenarios are obtained by rescaling the baseline FCC-ee statistical uncertainties by the luminosity ratio $\sqrt{\mathcal{L}_{\text{baseline}}/\mathcal{L}_{\text{scenario}}}$, using the luminosities reported in Tables~\ref{tab:descoped_runs} and~\ref{tab:upscoped_runs}. While a proper assessment of how systematic uncertainties would change in the descoped and upscoped FCC-ee scenarios requires dedicated simulations that are currently not available, several systematic uncertainties are expected to depend on the available integrated luminosity~\cite{Blondel:2024mry, Selvaggi:2025kmd, blondel_2025_3rj9z-94994}. In particular, this is the case for the asymmetries and ratios among the $Z$-pole EWPOs, as well as the Higgs production and decays measurements we consider. For these observables, when the dominant systematic uncertainty is statistically driven, we approximate its dependence on the integrated luminosity by rescaling it according to the same $1/\sqrt{\mathcal{L}}$ scaling used for statistical uncertainties. Otherwise, we keep the systematic uncertainty unchanged relative to the baseline FCC-ee scenario.

\paragraph{Theoretical predictions and uncertainties}
\label{subsec:theory_covmat}

We account for NLO QCD corrections in the SMEFT for most (HL)-LHC processes considered, as well as NLO electroweak corrections to $Zh$ production at the lepton colliders~\cite{Asteriadis:2024xts}.  Following the discussions in the ESPPU2026 Briefing Book~\cite{deBlas:2025gyz}, we include multiple sources of theory uncertainties on the SM predictions: those related to missing higher-order calculations,
those associated with the theoretical inputs
required for the experimental extraction of the EWPO measurements, and the parametric uncertainties arising from the propagation of experimental uncertainties on the measurements of input parameters. 
The first two sources of theory uncertainties are considered in this work through two benchmark scenarios: an ``aggressive'' scenario corresponding to substantial future improvements in theoretical predictions up to N$^4$LO, and an ``ideal'' scenario in which SM theory uncertainties are assumed to be negligible compared to experimental precision. The parametric uncertainties are of experimental origin and are always considered here, and we implemented them using the code from Ref.~\cite{Mildner:2024wbl}. Their size is affected by, first, the integrated luminosity as it affects the statistical uncertainty on the input parameters, and second, by a less precise determination of the top-quark mass in the absence of a top-quark run at $\sqrt{s}=365$ GeV. 
Similarly to previous SMEFiT studies, in this work we neglect theory uncertainties on the EFT predictions, but ensure that the Monte Carlo statistical uncertainties on the EFT predictions are kept below $1\%$. 
\\

The complete database hosting experimental projections and theoretical predictions is available in a public repository: \url{https://github.com/LHCfitNikhef/smefit_database}.
\section{Global SMEFT analysis} 
\label{sec:results}
\subsection{Comparison of the descoped scenarios}
\label{subsec:results_descoped}

\begin{figure}[h!]
    \centering
    \includegraphics[width=\linewidth]{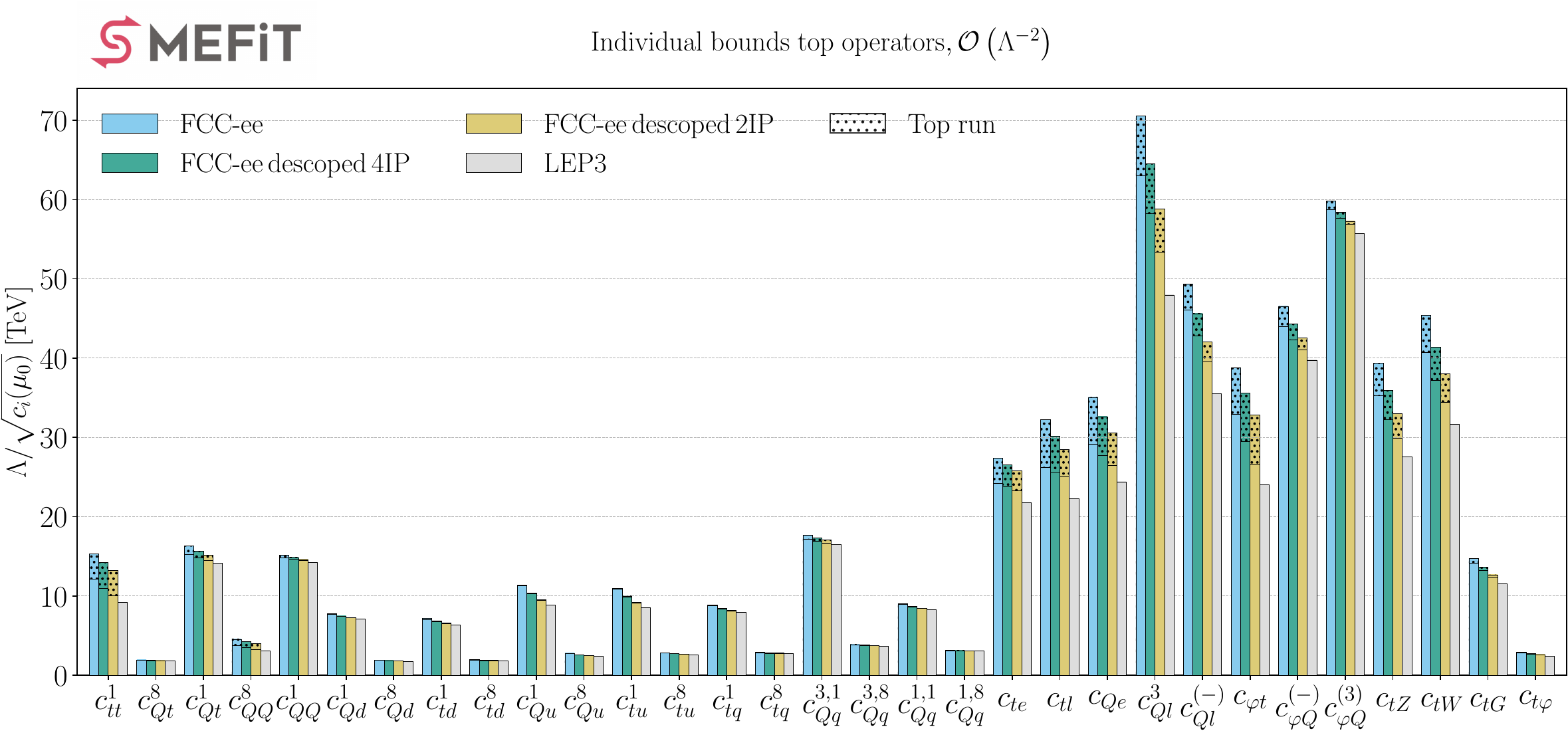}
    \includegraphics[width=\linewidth]{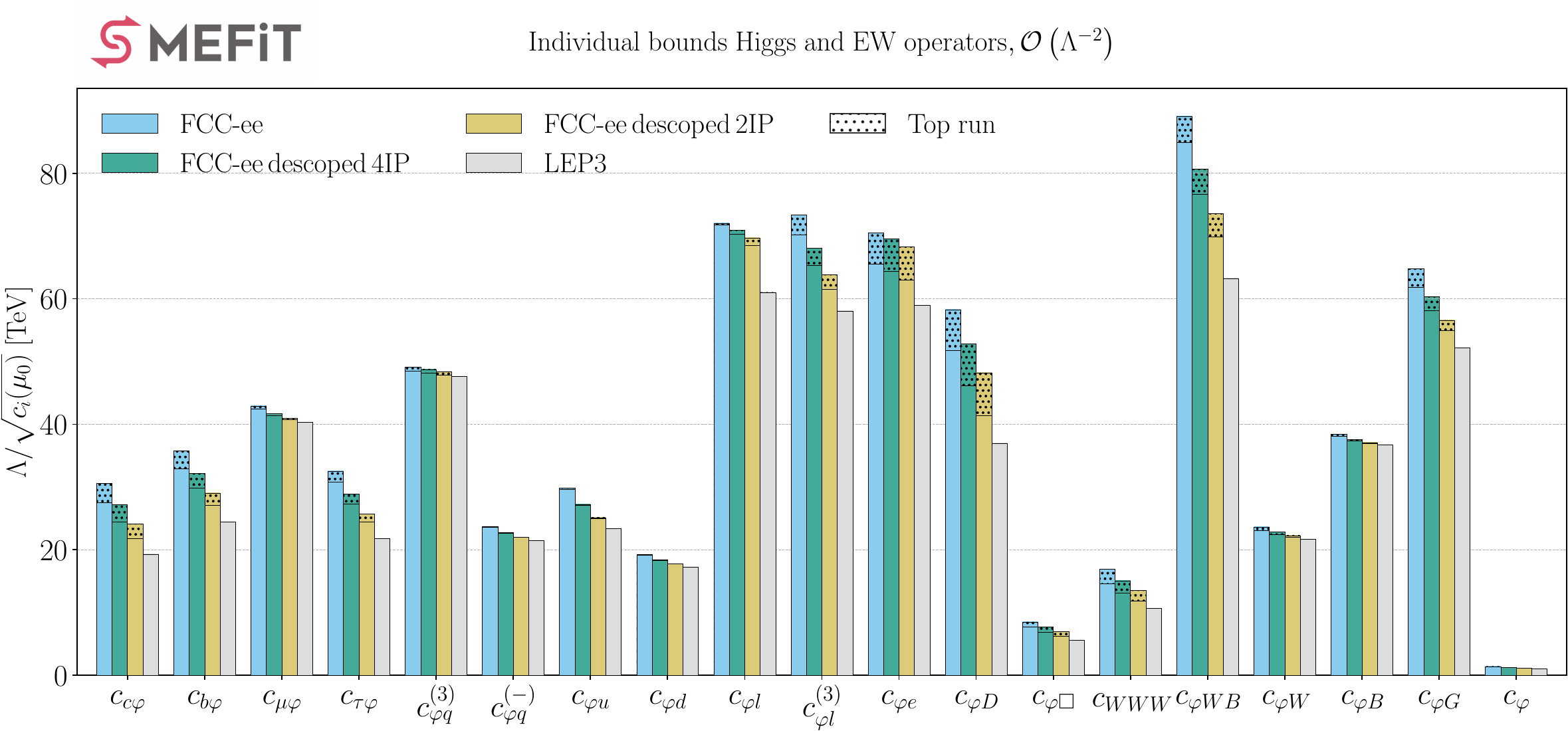}
    \caption{The 95\% C.I. on the mass reach of the top operators (top panel) and of the Higgs and electroweak operators (bottom panel) for the various descoping scenarios considered in this work. The results are obtained from individual fits at linear level in the EFT expansion. The dotted bars indicate the impact of the (staged) top-quark run. Aggressive theory uncertainties are adopted.
    }
    \label{fig:barplot-individual-aggressive}
\end{figure}

\begin{figure}[h!]
    \centering
    \includegraphics[width=\linewidth]{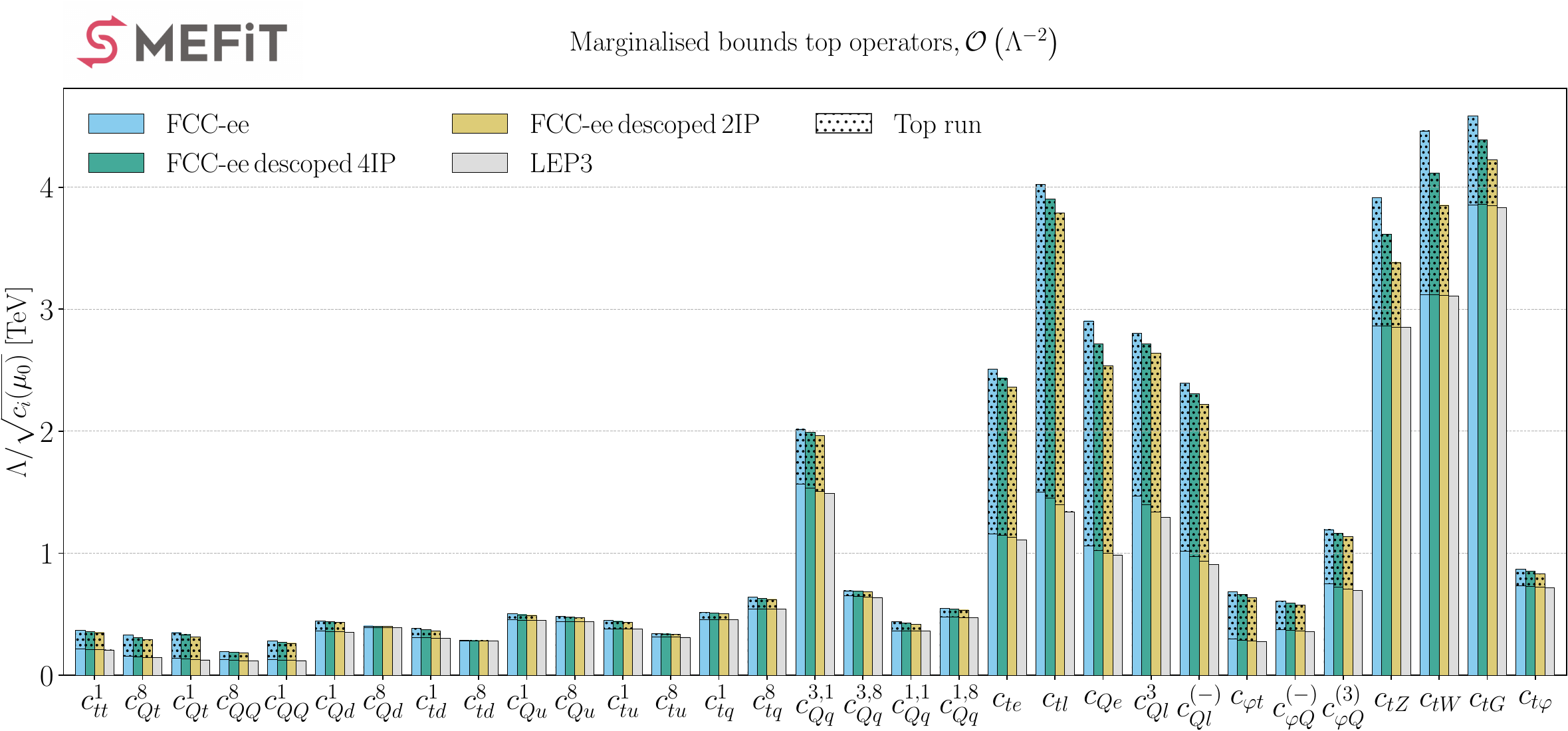}
    \includegraphics[width=\linewidth]{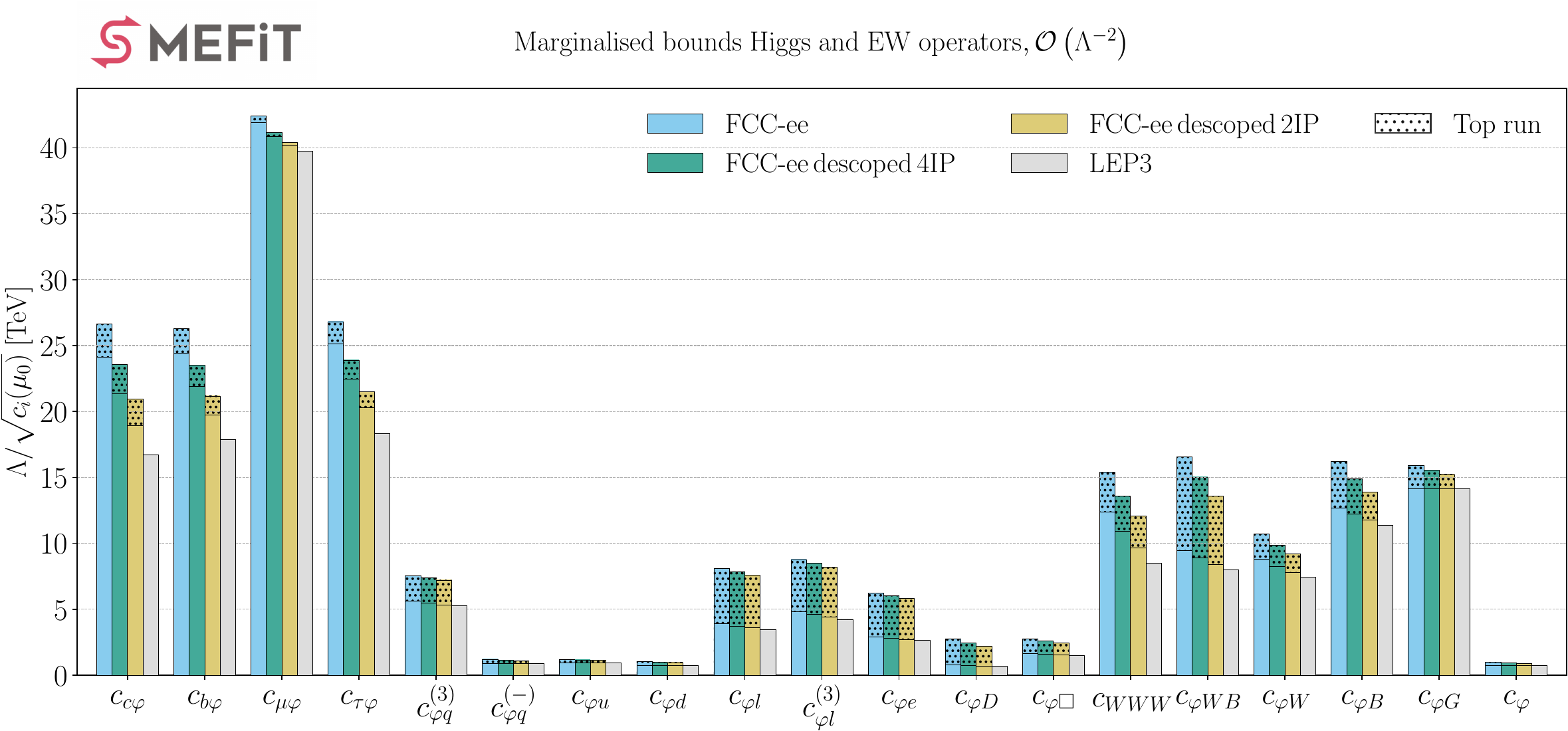}
    \caption{Same as Fig.~\ref{fig:barplot-individual-aggressive}, now for the marginalised fit where all operators are fitted simultaneously.}
    \label{fig:barplot-global-aggressive}
\end{figure}

\begin{figure}[h!]
    \centering
    \includegraphics[width=\linewidth]{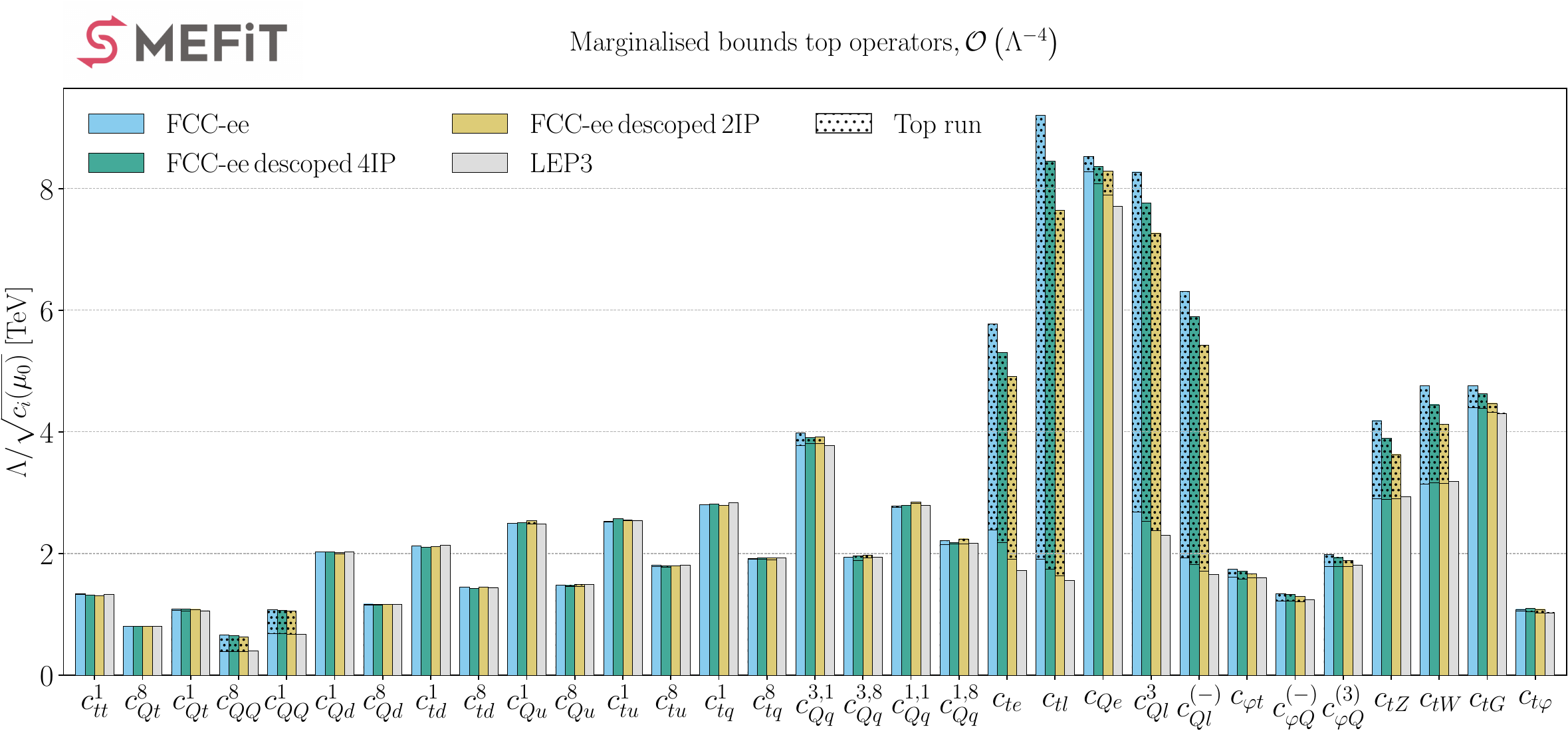}
    \includegraphics[width=\linewidth]{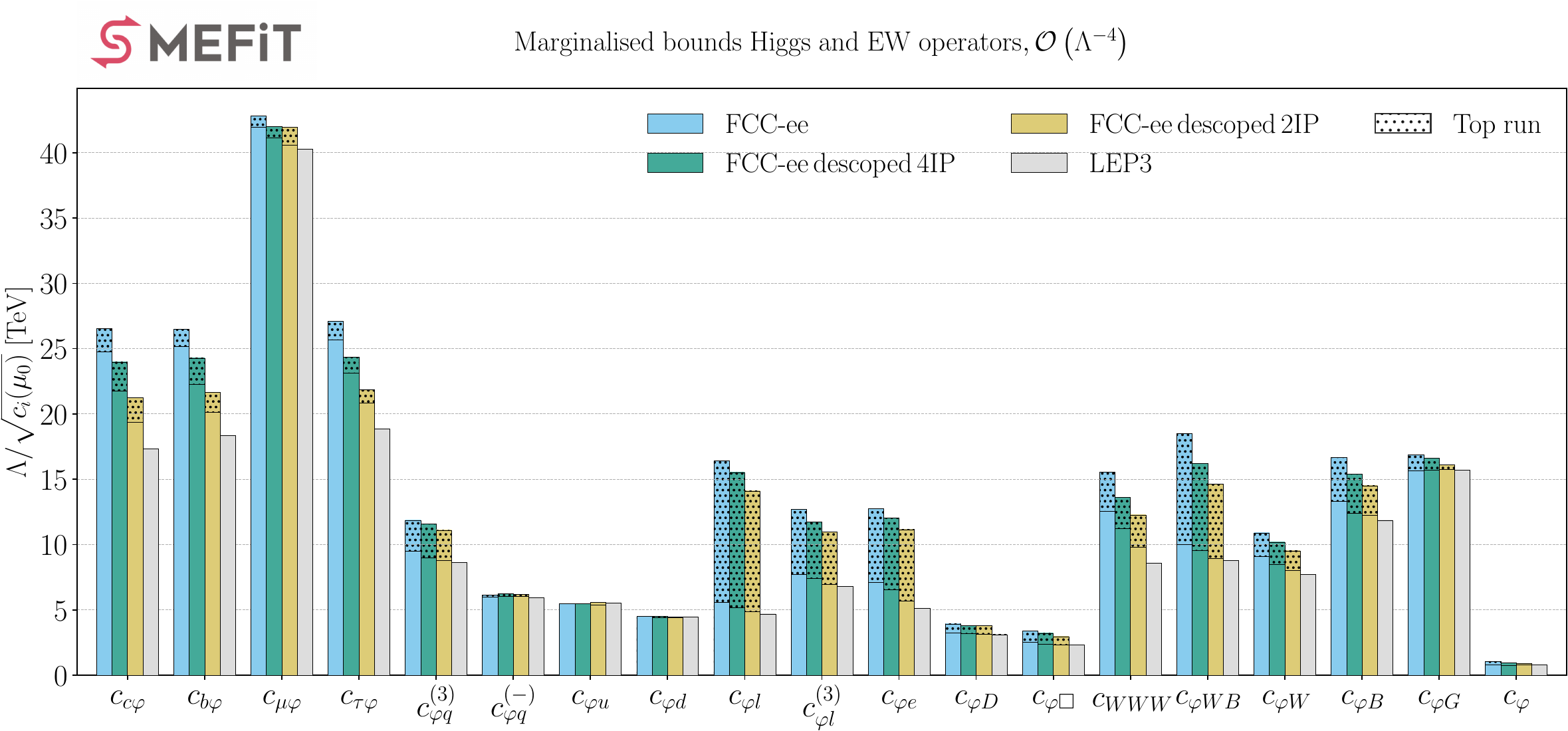}
    \caption{Same as Fig.~\ref{fig:barplot-global-aggressive}, now with quadratic EFT corrections taken into account.}
    \label{fig:barplot-global-aggressive-quadratic}
\end{figure}

In this section, we present the results of the SMEFT analysis for the descoped scenarios introduced in Sec.~\ref{subsec:descoped_scenarios}, and compare them to the baseline FCC-ee and LEP3 results. We present results for both individual fits, in which the Wilson coefficients are fitted one at a time, and a global fit, in which all 61 Wilson coefficients are fitted simultaneously and the constraint on each coefficient is obtained by marginalising over the remaining coefficients. For readability, we display here the results for 31 top coefficients and 19 Higgs and electroweak coefficients. The remaining coefficients corresponding to the 4-lepton and 2-lepton-2-light-quark operators, although taken into account in the marginalised fit, are not shown, as they do not provide additional information relevant to the arguments and conclusions presented here. For completeness, we present the corresponding bounds on these operators in App.~\ref{app:detailed_results}.

Figs.~\ref{fig:barplot-individual-aggressive} and
~\ref{fig:barplot-global-aggressive} compare, 
for the individual and global fits respectively, the 
new-physics scale reach  associated 
with the top, and Higgs and 
electroweak Wilson coefficients for the possible
descoped scenarios, the baseline FCC-ee, and LEP3. The results correspond to fits carried out at $\mathcal{O}(\Lambda^{-2})$ in the EFT expansion, and
are displayed in terms of bounds on $\Lambda/\sqrt{c_i(\mu_0)}$, which can be interpreted as 
the characteristic new physics mass scale being probed by the
SMEFT analysis assuming a coefficient $c_i(\mu_0)=1$. The aggressive theory uncertainty scenario is
adopted and the impact of the top-quark run is denoted by 
dotted bars.

\paragraph{Impact of reduced luminosity}
We first focus on the impact of different luminosities in the absence of the top-quark run. In Figs.~\ref{fig:barplot-individual-aggressive} and~\ref{fig:barplot-global-aggressive}, this is illustrated by the reach of the solid bars. In particular, we compare an FCC-ee with no top-quark run, a descoped FCC-ee with 2 IPs, and a descoped FCC-ee with 4 IPs. For completeness, we also present the projected performance of LEP3.
Starting with the individual fits in Fig.~\ref{fig:barplot-individual-aggressive}, the same performance pattern is observed overall: FCC-ee with no top-quark run dominates, followed by the 4-IP and 2-IP descoped scenarios, and finally LEP3. In the descoped scenarios, the effects of the reduced luminosity are particularly noticeable for some two-lepton-two-heavy quark operators, the Yukawa couplings, the top dipole operators $c_{tW}$, $c_{tZ}$ and $c_{tG}$, and the purely bosonic operators. Taking the specific example of $c_{\varphi W B}$, we find that while a LEP3-only scenario has a reach of just above 60 TeV, a full-luminosity FCC-ee scenario without the top run reaches just above 80 TeV, corresponding to approximately a $30\%$ improvement. The four-quark operators do not contribute to any of the FCC-ee observables considered here at tree level and are therefore probed only through renormalisation group (RG) effects. Considering the one-loop solution to the RG equations, the colour-singlet operators and the colour-octet operator $c_{QQ}^8$ already generate, at leading-logarithmic order, Wilson coefficients that enter the EWPOs at tree level. By contrast, the remaining colour-octet operators generate such coefficients only at higher orders in the logarithmic expansion, which explains their lower reach on the new-physics scale.

Turning next to the marginalised results displayed in Fig.~\ref{fig:barplot-global-aggressive}, we first note that the overall reach of the operators is drastically reduced compared to the individual fits, particularly in the top sector where the reduction factor is often around 20. This is an expected behaviour showing that several operators are correlated in the simultaneous fit, increasing the uncertainty on each of them separately.

Comparing our different descoped scenarios, we observe that the impact of the reduced luminosity is milder than in the individual fits, suggesting that the dominant source of uncertainties in the global analysis is not only statistically driven, a point which we address in detail in Sec.~\ref{sec:operator_subset}. The Yukawa operators are an exception to this statement, as their behaviour and reach did not change much compared to the individual fits, suggesting that they do not correlate much with other operators. The other operators tend to change very little between the scenarios, as illustrated by $c_{\varphi W B}$, which goes, as we previously pointed out, from a 30\% difference between FCC-ee and LEP3 in the individual case to only a few percent in the marginalised fit. Overall, we note that the differences between the various scenarios are very moderate when the top-quark run is not included, and no variant of the FCC-ee truly outperforms LEP3.

For completeness, we provide in Fig.~\ref{fig:barplot-global-aggressive-quadratic} the marginalised results obtained in the presence of quadratic EFT corrections. Comparing it to its linear counterpart shown in Fig.~\ref{fig:barplot-global-aggressive}, we observe that the mass reach increases by at least a factor of two for the four-fermion operators due to their large quadratic corrections, while the dipole operators instead show a similar reach in both. This is especially visible for the two-lepton-two-heavy-quark operators, where the mass reach increases from around 4 TeV in $c_{t\ell}$ to a little more than 8 TeV once quadratic corrections are turned on. In this context, it also worth pointing out that the relative contribution to the bounds coming from the top-quark run can either increase between linear and quadratic fits, as for $c_{t\ell}$, or decrease, as for $c_{Qe}$ and some of the 4-heavy operators like $c_{tt}^1$. Whether the relative contribution from the top-quark run increases or decreases depends in general on whether that specific operator is well aligned with respect to the principal directions of the posterior covariance matrix $C$ of EFT coefficients. However, a basis independent statement can be made by considering the posterior volume corresponding to $C$, defined as
\begin{equation}
    V_{\rm post} \equiv \log \sqrt{\det C} \, ,
\end{equation}
as a measure of the overall constraining power. We have verified that $V_{\rm post}$ increases by 8\% once the top-quark run is excluded from the baseline FCC-ee scenario in the quadratic EFT fit, indicating that overall the top-quark run brings in additional constraining power.

The individual bounds obtained when quadratic EFT corrections are included differ only marginally from their linear counterparts shown in Fig.~\ref{fig:barplot-individual-aggressive}. They are therefore not presented here, but are included in App.~\ref{app:detailed_results} for completeness. For individual fits, the region of EFT space probed at FCC-ee and LEP3 is dominated by linear corrections, making quadratic terms irrelevant. This contrasts with the global EFT fit, where quadratic contributions can help reduce correlations and therefore have a more significant impact, as also pointed out in Ref.~\cite{Armadillo:2026mvp}. In the following, unless specified, we focus on the linear EFT results.

\paragraph{Impact of the top-quark run} 

As discussed above, excluding a run at $\sqrt{s}=365$ GeV does not lead to large differences among the four scenarios for most of the Wilson coefficients considered. The inclusion of a staged top-quark run in the FCC-ee scenarios is the key feature that allows them to perform better than LEP3, especially in the marginalised analysis, as depicted in Figs.~\ref{fig:barplot-individual-aggressive} and~\ref{fig:barplot-global-aggressive}, where the impact of a top-quark run is represented by a dotted bar. For the descoped scenarios, a staged top-quark run corresponds to a run at $\sqrt{s}=365$ GeV with integrated luminosity equal to 0.365 (0.6) times that of the top-quark run at the baseline FCC-ee, for the 2 (4) IP descoped scenario. In the individual fit shown in Fig.~\ref{fig:barplot-individual-aggressive}, we see that the reach increases are typically under 10\%. One of the largest increases affects $c_{tl}$, moving its reach from about 25 TeV to slightly above 30 TeV, corresponding to a $\mathcal{O}(20\%)$ difference. On the other hand, in the marginalised fit shown in Fig.~\ref{fig:barplot-global-aggressive}, we observe a strong impact of the top-quark run, with large reach increases, sometimes of the order of, or greater than, 100\%. For example, $c_{tl}$ moves from 1.5 TeV to 4 TeV in the baseline FCC-ee scenario.
As discussed in Ref.~\cite{Armadillo:2026mvp}, in the individual fits, most of the constraining power of the FCC-ee comes from the unprecedented precision at the $Z$-pole, to which most of the Wilson coefficients run through RG evolution. However, in the marginalised fit, coefficients are subject to correlations that result in a relative weakening of the constraints compared to the individual fit results. The top-quark run provides additional observables with new operator dependencies that help break these correlations, resulting in relatively higher mass reaches compared to the corresponding improvement in the individual fit, as is apparent in the two-lepton-two-heavy quark operators that enter $t\bar{t}$ production at tree level.

\begin{figure}[h!]
    \centering
    \includegraphics[width=0.49\linewidth]{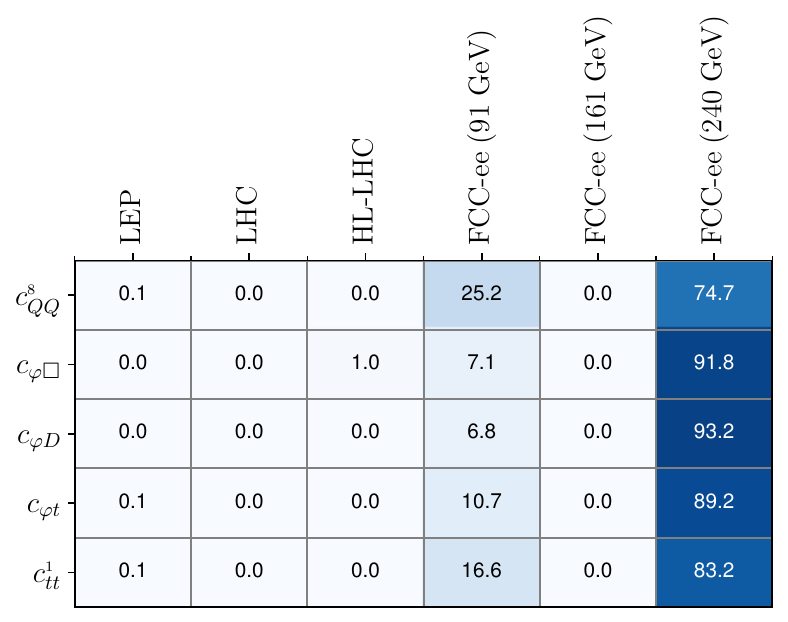}
\includegraphics[width=0.49\linewidth]{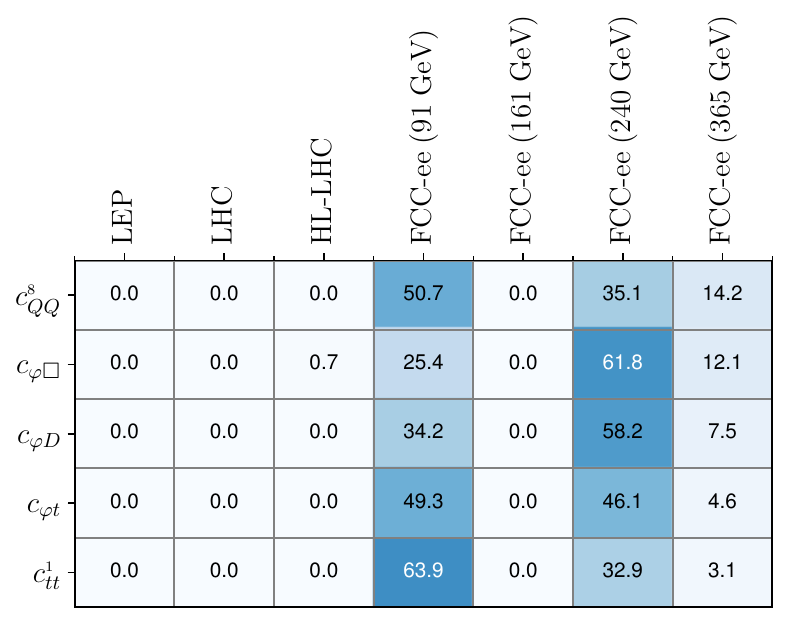}
    \caption{Comparison of diagonal entries of the Fisher Information Matrix for 
    a selected subset of operators in the absence (left) and in the presence (right) of a top-quark run in the baseline FCC-ee scenario. 
    The top-quark run drastically reduces the parametric uncertainties associated with 
    the top-quark mass, indirectly increasing the sensitivity of the $Z$-pole observables. Aggressive theory uncertainties are adopted. 
    }
    \label{fig:reduced_fisher}
\end{figure}

A further consequence of the top-quark run is the reduction of parametric uncertainties, particularly those associated with the top-quark mass $m_t$, which is expected to be measured within $\mathcal{O}(10)$ MeV precision at future colliders at the $t \bar t$ threshold and enhances the constraining power of the $Z$-pole  \cite{FCC:2025lpp}. To explicitly see this, we show in Fig.~\ref{fig:reduced_fisher} the diagonal elements of the Fisher information matrix for some of the affected coefficients ($c_{QQ}^8, c_{\varphi \Box}, c_{\varphi D}, c_{\varphi t}, c_{tt}^1$), with and without the top-quark run. The Fisher information matrix is defined at linear order in the EFT expansion as \cite{Ethier:2021bye}
\begin{equation}
    F_{ii'} = \sum_{j, j'=1}^{n_{\rm dat}}\kappa_{i,j}\Sigma_{j,j'}^{-1}\kappa_{i',j'} \, ,
    \label{eq:fisher_0}
\end{equation}
where the sum runs over all data points that enter the fit, and $\kappa_{i,j}$ denotes the linear EFT correction to data point $j$ induced by the EFT parameter $i$, while $\Sigma_{ij}$ represents the total fit covariance matrix including both theoretical and experimental uncertainties. To identify which classes of measurements constrain the different directions
in the EFT parameter space, we decompose Eq.~\eqref{eq:fisher_0} into
contributions from different dataset groups, \begin{equation}
    F = \sum_g F^{(g)} \, .
    \label{eq:fisher_grouping}
\end{equation}
where $g$ denotes groups of datasets composed exclusively of either LEP, LHC, HL-LHC, or FCC-ee data at separate centre of mass energies.
Inspecting Fig.~\ref{fig:reduced_fisher}, we find that, when the top-quark run is absent, these coefficients are primarily constrained from the run at $\sqrt{s} = 240$ GeV, while once the top-quark run is added and the uncertainty on $m_t$ is consequently reduced, the constraining power of the $Z$-pole increases.

\begin{figure}[htbp]
    \centering
    \includegraphics[width=0.7\linewidth]{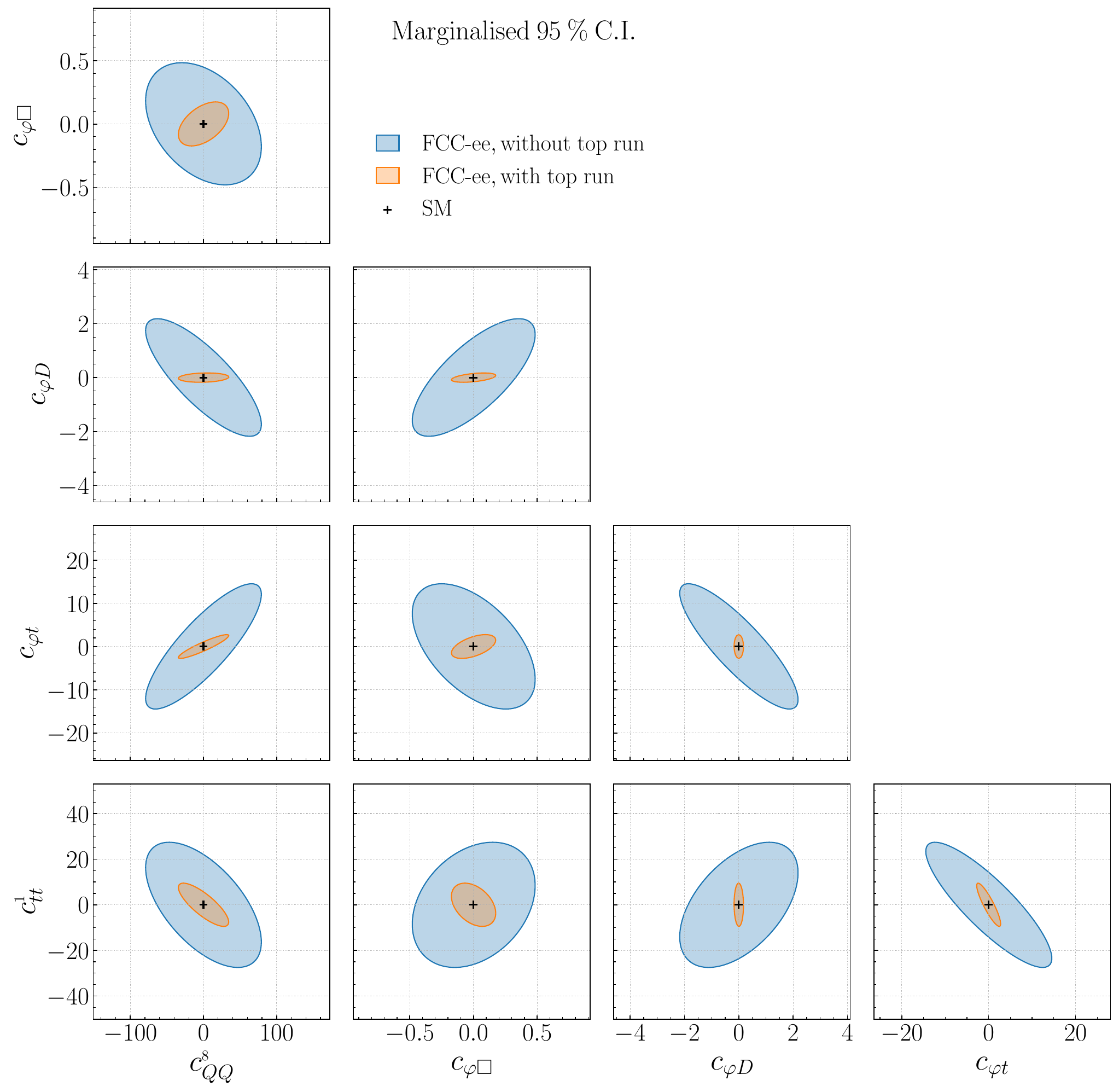}
    \caption{95 \% C.I. for the subset of operators $c_{tt}^1$, $c_{QQ}^8$, $c_{\varphi \Box}$, $c_{\varphi D}$ and $c_{\varphi t}$ entering the 61-dimensional linear global SMEFT fit. We compare the outcome of fits to the baseline FCC-ee scenario with and without a top-quark run, while always including aggressive theory uncertainties. 
    }
   \label{fig:top_impact_cornerplot}
\end{figure}

These effects are further illustrated in Fig.~\ref{fig:top_impact_cornerplot}, which compares the impact on the 95\% C.I bounds for the same representative subset of operators in the global FCC-ee fit with and without top-quark run. We fit all 61 coefficients and marginalise over 59 of them to obtain the 2-dimensional bounds displayed. We find that the reduction of the bounds in the presence of the top-quark run comes primarily from a breaking of the correlations between the different operators. This is driven by the new observables added to the fit, which probe complementary directions, as well as reduced parametric uncertainties in the EWPOs due to a more precise determination of the top-quark mass. 

\begin{figure}[htbp]
    \centering
    \includegraphics[width=0.7\linewidth]{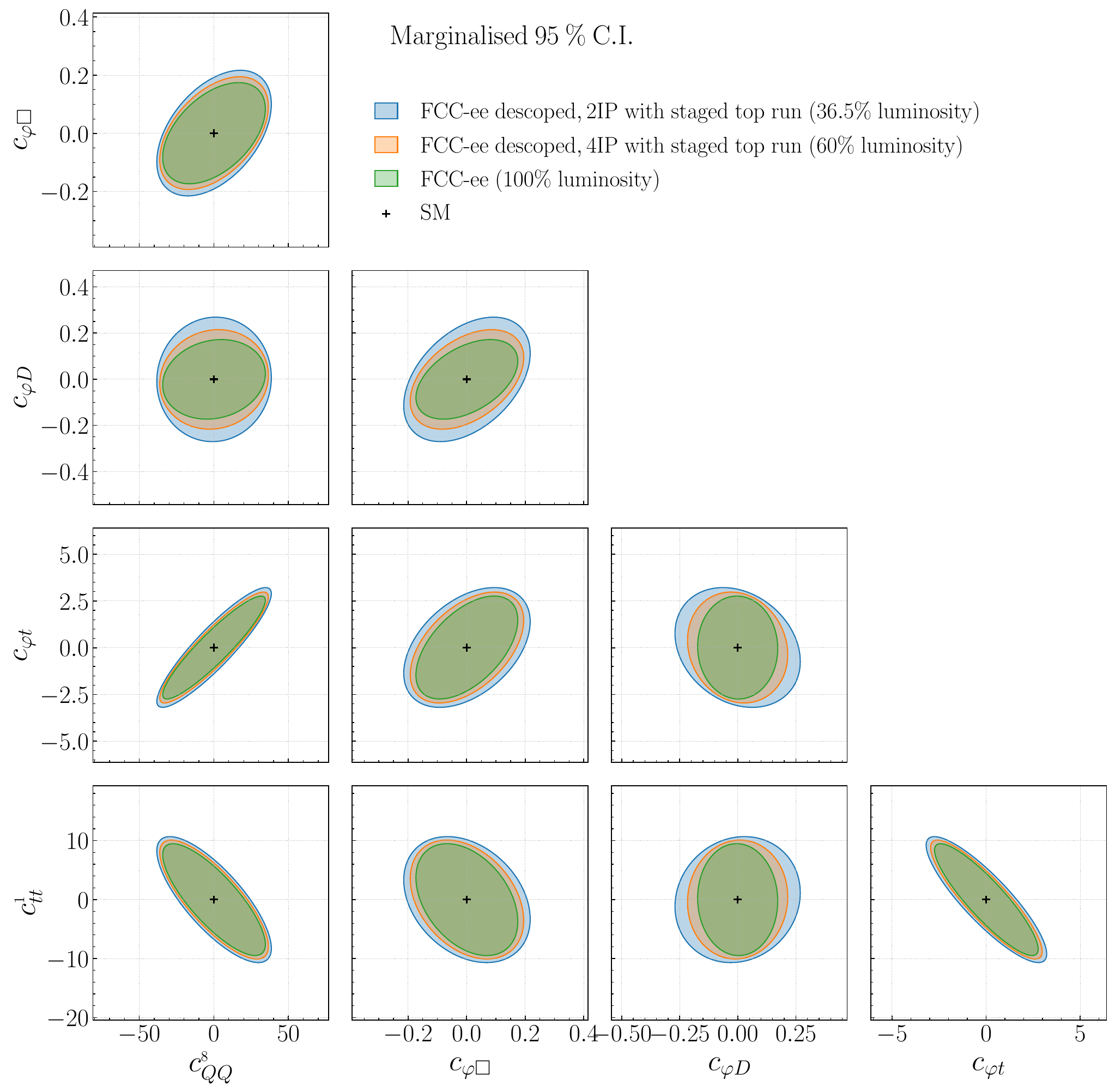}
    \caption{Same setup as Fig.~\ref{fig:top_impact_cornerplot}, now comparing the impact of different FCC-ee descoping scenarios. We compare the outcome of the baseline FCC-ee scenario to the 2 and 4IP descoped scenarios with a staged top-quark run.
    }
\label{fig:lumi_scaling_cornerplot}
\end{figure}

While the top-quark run markedly improves the mass reaches for the different FCC-ee scenarios, the effect of the reduced integrated luminosity in the descoped scenarios becomes relatively modest once a staged top-quark run is included, as shown in Figs.~\ref{fig:barplot-individual-aggressive} and~\ref{fig:barplot-global-aggressive}. To further compare the impact of the reduced luminosity with that of including or omitting the top-quark run, we show in Fig.~\ref{fig:lumi_scaling_cornerplot}, the counterparts of the corner plots in Fig.~\ref{fig:top_impact_cornerplot}, for the baseline FCC-ee and the 2- and 4-IP scenarios, all with a top-quark run included. Fig.~\ref{fig:lumi_scaling_cornerplot} shows that the luminosity reduction does not substantially affect the constraints, leading only to modest changes in the sizes of the ellipses. Even a factor of three difference in luminosity has a considerably smaller impact than the inclusion or omission of the top-quark run, which helps to break correlations between the coefficients, as shown in Fig.~\ref{fig:top_impact_cornerplot}. This suggests that at a descoped FCC-ee, the inclusion of a staged top-quark run should be favoured over attempting to match
the target luminosity of the baseline FCC-ee in the lower energy runs.

\begin{figure}
    \centering
    \includegraphics[width=\linewidth]{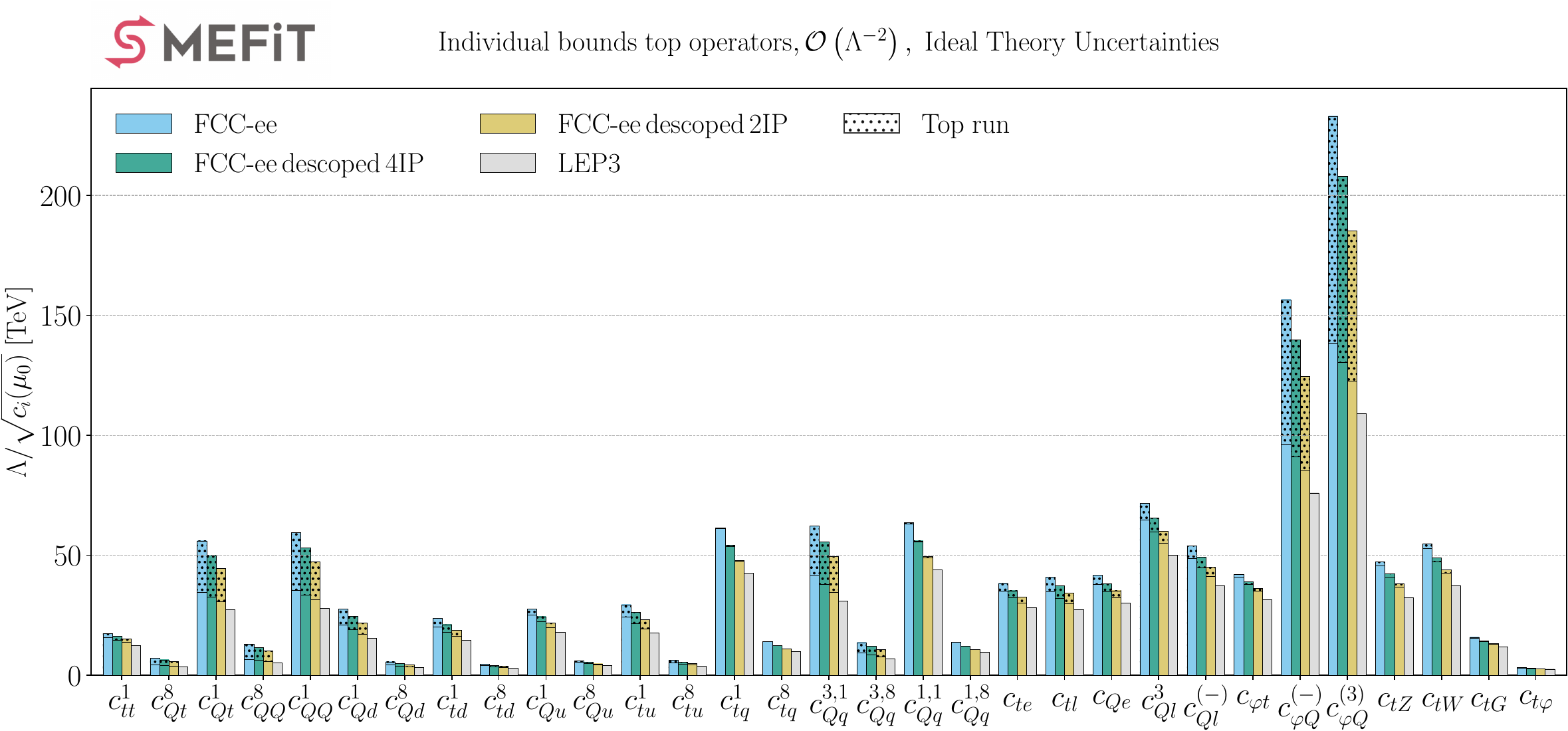}
    \includegraphics[width=\linewidth]{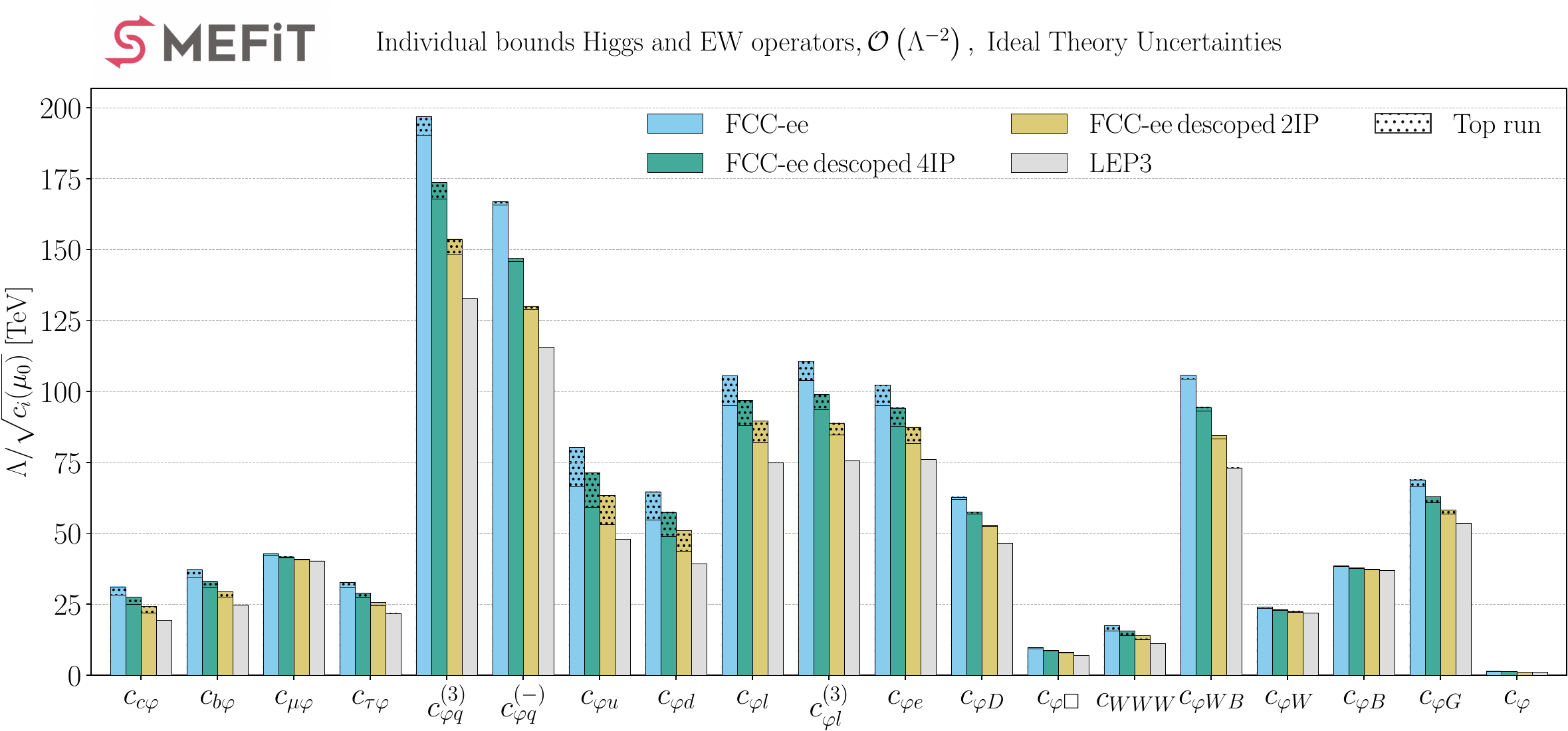}
    \caption{The 95\% C.I. on the mass reach of the top (top panel) and Higgs and electroweak operators (bottom panel) in the individual fit for the various descoping scenarios considered in this work. The dotted bars indicate the impact of the (staged) top-quark run. The ideal theory scenario corresponding to the absence of theory uncertainties except for the parametric ones is adopted.}
    \label{fig:ideal_ind}
\end{figure}

\paragraph{Impact of theory uncertainties} We now turn our attention to the impact that theory uncertainties have on the SMEFT parameter space. To this aim, Fig.~\ref{fig:ideal_ind} shows the same results as the individual fits in Fig.~\ref{fig:barplot-individual-aggressive}, but in the ideal theory scenario, where theory uncertainties are taken to be negligible, except for parametric uncertainties which are kept. Comparing side by side Fig.~\ref{fig:ideal_ind} with the results in the aggressive theory scenario, we find that for the individual fit, even in the aggressive scenario the theory uncertainties are still very large and removing them leads to a significant improvement of the mass reaches at all collider scenarios. Two striking examples are $c_{\varphi Q}^{(3)}$ and  $c_{\varphi Q}^{(-)}$, whose reaches go from around 60 and 45 TeV respectively in the aggressive theory scenario to around 100-240 TeV and 75-150 TeV depending on the collider scenarios when the theory uncertainties are neglected. For the Higgs and electroweak operators, the impact is also important with many operators improving their mass reach by 50\% and up to factors of 5-7, for example $c_{\varphi q}^{(3)}, c_{\varphi q}^{(-)}, c_{\varphi u}, c_{\varphi d}, c_{\varphi e}$.
Furthermore, as expected, the differences between the collider scenarios become more pronounced when theory uncertainties are negligible, highlighting that significant advances in theory calculations are required to fully exploit the large luminosities of the FCC-ee programme. This is in particular the case for the colour-singlet four-quark operators, as well as some of the operators entering the EWPOs such as $c_{\varphi Q}^{(3)}, c_{\varphi Q}^{(-)}, c_{\varphi u}, c_{\varphi d}$. 

These observations are consistent with Ref.~\cite{Armadillo:2026mvp}, where theory uncertainties were found to have a large impact in the individual fit and a much more moderate impact in the marginalised one. This can be understood as follows. As observed in Ref.~\cite{Armadillo:2026mvp}, in the individual fits, a substantial fraction of the constraining power of the FCC-ee, and hence of the descoped scenarios, comes from the unprecedented precision of the $Z$-pole observables, to which most operators flow via RGE. Moreover, the theory uncertainties introduced in Sec.~\ref{sec:analysis} affect most strongly the $Z$-pole observables, where they enter in the form of missing higher-order corrections, background subtraction in the EWPO measurements, and parametric uncertainties. In the individual fits of Fig.~\ref{fig:ideal_ind}, the improvement in the mass reaches is primarily driven by the $Z$-pole run, whose constraining power increases further once theory uncertainties are neglected. This is not the case for the marginalised fit, where as discussed above the indirect constraints from the $Z$-pole observables mix with the correlations among the coefficients such that the constraints from the $Z$-pole run do not dominate the fit anymore. We show in App.~\ref{app:detailed_results} in Fig.~\ref{fig:ideal_glob} the marginalised fit results in the ideal theory scenario, where only a small improvement is observed compared to the aggressive theory scenario. We address this further in Sec.~\ref{subsec:HLLHC_bottleneck}.

\subsection{Impact of an upscoped run}
\label{subsec:results_upscoped}
\begin{figure}[htbp]
    \centering
    \includegraphics[width=\linewidth]{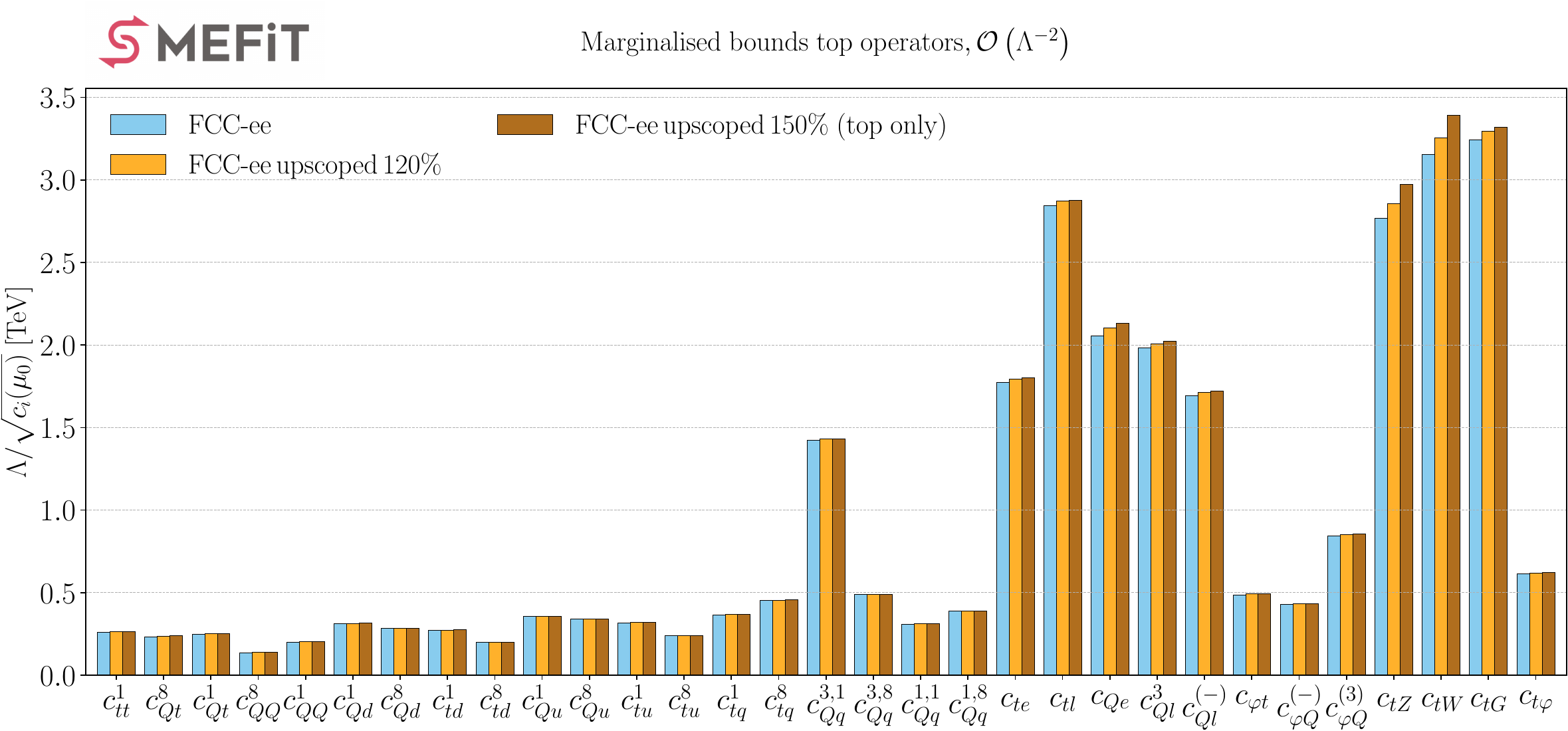}
    \includegraphics[width=\linewidth]{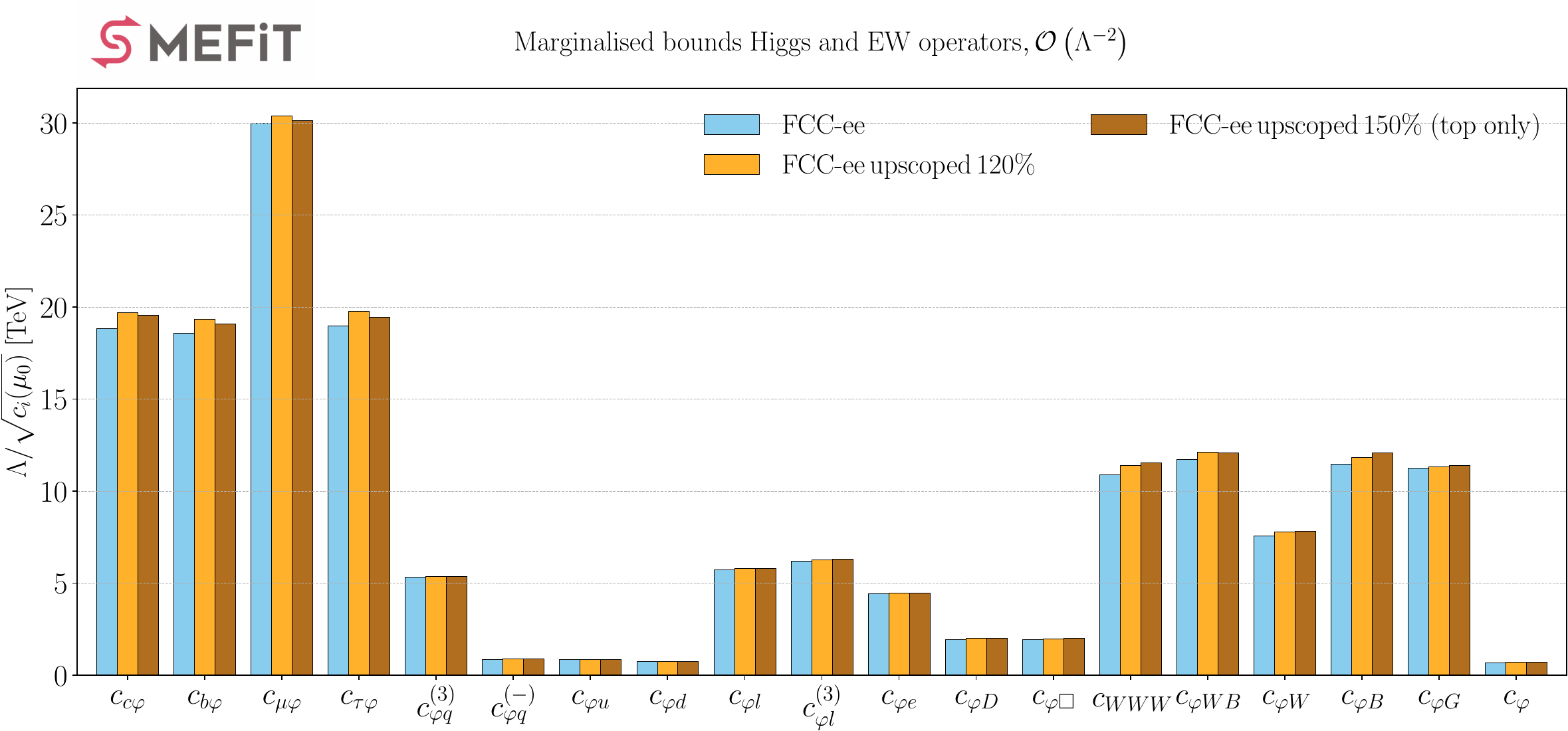}
    \caption{The 95\% C.I. on the mass reach for the FCC-ee upscoped scenarios in the global marginalised fit for top operators (upper panel) and Higgs and electroweak operators (bottom panel) at the linear level in the EFT expansion. We display the 120\% upscoped scenario where the luminosity is increased by 20\% across all energy runs, and a targeted 150\% upscoped scenario where the luminosity is increased by 50\% only for the top-quark run while keeping the same luminosities for the lower energy runs as the baseline FCC-ee, which we also display for clarity. Aggressive theory uncertainties are adopted.} 
    \label{fig:barplot-global-upscoped}
\end{figure}

We now explore two upscoped FCC-ee scenarios, one where all the runs get an additional $20 \%$ integrated luminosity compared to the baseline FCC-ee, and one where only the top-quark run at $\sqrt{s}= 365$ GeV is modified and receives $50\%$ additional luminosity. The results are shown in Fig.~\ref{fig:barplot-global-upscoped} for the global marginalised fit at linear order in the EFT expansion for the top operators (upper panel) and Higgs and electroweak operators (lower panel) separately. The equivalent results for individual fits are shown in App.~\ref{app:detailed_results} in Fig.~\ref{fig:barplot-individual-upscoped}.

Inspecting Fig.~\ref{fig:barplot-global-upscoped}, we find that the overall impact of the additional luminosity is marginal, which indicates that the projected uncertainties on the measurements are not statistically dominated, to which we will return in Sec.~\ref{sec:operator_subset}. We further note that no single scenario consistently dominates across the board: the upscoped top-quark run does not always outperform or underperform the overall upscoped FCC-ee scenario. However, most coefficients benefit more from the overall luminosity rescaling to $120\%$, as this provides a larger integrated luminosity also for the $Z$-pole run, which sets the leading constraints in the majority of cases \cite{Armadillo:2026mvp, Allwicher:2024sso}. The upscoped top-quark run is more useful to further constrain the top dipole operators, and provides competitive bounds compared to the $120\%$ upscoped scenario for some of the operators entering the EWPOs through reduced parametric uncertainties, such as $c_{\varphi W B}$ and $c_{\varphi l}^{(3)}$. 

Overall, Fig.~\ref{fig:barplot-global-upscoped} further indicates that, once a top-quark run is included, the exact luminosity of that run has a comparatively limited effect on the global SMEFT fit. In contrast, omitting the top-quark run leads to substantially stronger changes in the bounds, as shown in Fig.~\ref{fig:top_impact_cornerplot}.

\subsection{Mapping out the fit sensitivity}
\label{sec:operator_subset}

As observed in Sec.~\ref{subsec:results_descoped} and \ref{subsec:results_upscoped}, the various upscoped and descoped scenarios for the FCC-ee do not lead to large variations in mass reaches once a top-quark run is considered, particularly for global fits. This suggests that the projected FCC-ee statistical uncertainties are not the limiting factor. The same is true of the parametric uncertainties of the $Z$-pole observables determined from the projected FCC-ee top-quark run. While the difference between an FCC-ee \emph{with or without} a top-quark run is important, as shown in Fig.~\ref{fig:top_impact_cornerplot}, the comparison between a \emph{targeted upscoped} and a \emph{nominal} top-quark run yields only marginal differences. This suggests that the uncertainty bottleneck lies elsewhere, and identifying it is the focus of this section.

\paragraph{Breaking flat directions with HL-LHC}
\label{subsec:HLLHC_bottleneck}
 We focus first on the global marginalised fits. A detailed exploration reveals that the main limitation comes from the systematic uncertainties associated with the LHC and HL-LHC observables. This statement may appear counter-intuitive in view of the Fisher Information Matrix associated with the fits, such as the reduced version shown in Fig.~\ref{fig:reduced_fisher}. 
At first sight, the projected FCC-ee measurements appear to completely dominate those from the LHC and HL-LHC to such an extent that the latter provide little additional constraining power. However, the Fisher information matrix shown in Fig.~\ref{fig:reduced_fisher} only carries the information corresponding to a linear individual fit. When the 61 dimensions of the marginalised fits are constrained simultaneously, the parameter space 
suffers from degeneracies with respect to the FCC-ee-only likelihood. These 
flat directions require the inputs from the (HL)-LHC in order to be broken. Consequently, the (HL)-LHC measurements become the bottleneck in the global fits, explaining the limited impact of the proposed upscoped FCC-ee scenarios.

\begin{figure}[ht!]
    \centering
    \includegraphics[width=0.7\linewidth]{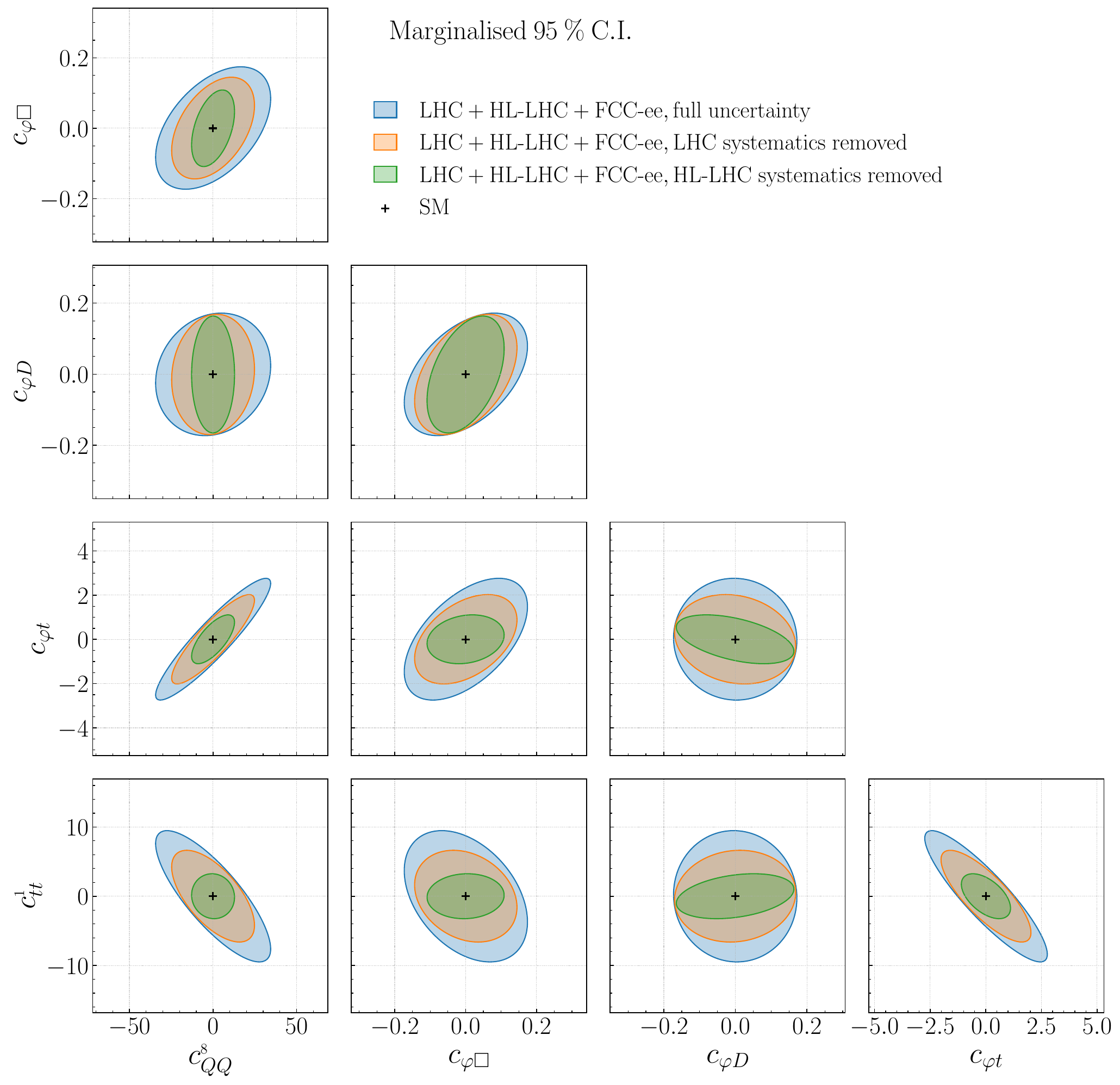}
    \caption{95 \% C.I. for the subset of operators $c_{tt}^1$, $c_{QQ}^8$, $c_{\varphi \Box}$, $c_{\varphi D}$ and $c_{\varphi t}$ entering the global linear SMEFT fit. Starting from the baseline FCC-ee, we compare the impact of removing the LHC systematics or the HL-LHC ones.}
    \label{fig:LHC_syst_corner_plot}
\end{figure}

To visualise this, we show in Fig.~\ref{fig:LHC_syst_corner_plot}, a
reduced corner plot for the baseline FCC-ee scenario, comparing different treatments of the systematic uncertainties associated with the LHC and HL-LHC observables. In this plot, we assume the aggressive theory scenario, in which theory uncertainties are neglected except for the parametric ones. We consider four cases (but only plot the first three for readability) for the treatment of systematic uncertainties: one using the full covariance matrix, one in which only the systematic uncertainties of the LHC observables are removed, one in which only those of the HL-LHC projections are removed, and finally one in which both the LHC and HL-LHC systematic uncertainties are removed. We observe that removing the systematics from the existing LHC measurements alone already greatly reduces the SMEFT bounds, with an even larger effect when the HL-LHC systematics are removed instead. Furthermore, removing the HL-LHC 
systematics alone yields results that are visually indistinguishable from removing them from both LHC and HL-LHC, which is why this fourth case is not represented in Fig.~\ref{fig:LHC_syst_corner_plot}. This suggests that, as expected, the HL-LHC projections dominate 
the fit over the LHC measurements. From this observation, while new experimental measurements from LHC Run 3 could still interplay with future FCC-ee measurements, it may be sufficient to focus solely on the HL-LHC projections to capture the same benefits.
The key point is that the projected FCC-ee measurements rely on input from the LHC, particularly from the HL-LHC run, to disentangle the global SMEFT fit. This remains true regardless of the assumptions made about experimental uncertainties at the FCC-ee. The LHC should therefore be regarded as a fundamentally complementary programme to the FCC-ee, rather than one that will be rendered obsolete by it.

We show in Fig.~\ref{fig:LHC_syst_corner_plot_quadratic} the version of Fig.~\ref{fig:LHC_syst_corner_plot} including the quadratic SMEFT corrections. We observe that the impact of removing the LHC systematic uncertainties has been lessened compared to the linear-only case. This is due to the fact that some correlations between operators have already been broken by the inclusion of the quadratic corrections, for instance between $c^1_{tt}$ and $c^8_{QQ}$. However, we still observe a much greater change than the one caused by the variation of the FCC-ee luminosity. Therefore, the observed synergy between the LHC and the FCC-ee is preserved in the quadratic case.

Focusing again on the linear-only fit, in order to identify which classes of measurements constrain the different directions
in the EFT parameter space, we diagonalise the Fisher information matrix from Eq.~\eqref{eq:fisher_0}, 
\begin{equation}
    F = V\,\mathrm{diag}(\lambda_1,\ldots,\lambda_n)\,V^{\!\top} \, ,
\end{equation}
to obtain an orthonormal basis of eigenmodes $v_k$, ordered by decreasing Fisher
information $\lambda_k$. Recalling Eq.~\eqref{eq:fisher_grouping}, the contribution of dataset group $g$ to
eigenmode $k$ can then be obtained by projecting $F^{(g)}$ onto the corresponding
eigenmode,
\begin{equation}
    \lambda_k^{(g)} = v_k^{\!\top} F^{(g)} v_k \, .
\end{equation}
By linearity of the Fisher matrix we have
\begin{equation}
    \sum_g \lambda_k^{(g)}
    = v_k^{\!\top} F v_k
    = \lambda_k \, ,
\end{equation}
so that the fractional contribution of each group is simply $ \lambda_k^{(g)}/\lambda_k $, which sum to unity by construction. This fraction can be interpreted as the relative sensitivity of the datasets in group $g$ to the C.I. bounds along principal direction $v_k$ in the global EFT fit. 

\begin{figure}[h!]
    \centering
    \includegraphics[width=\linewidth]{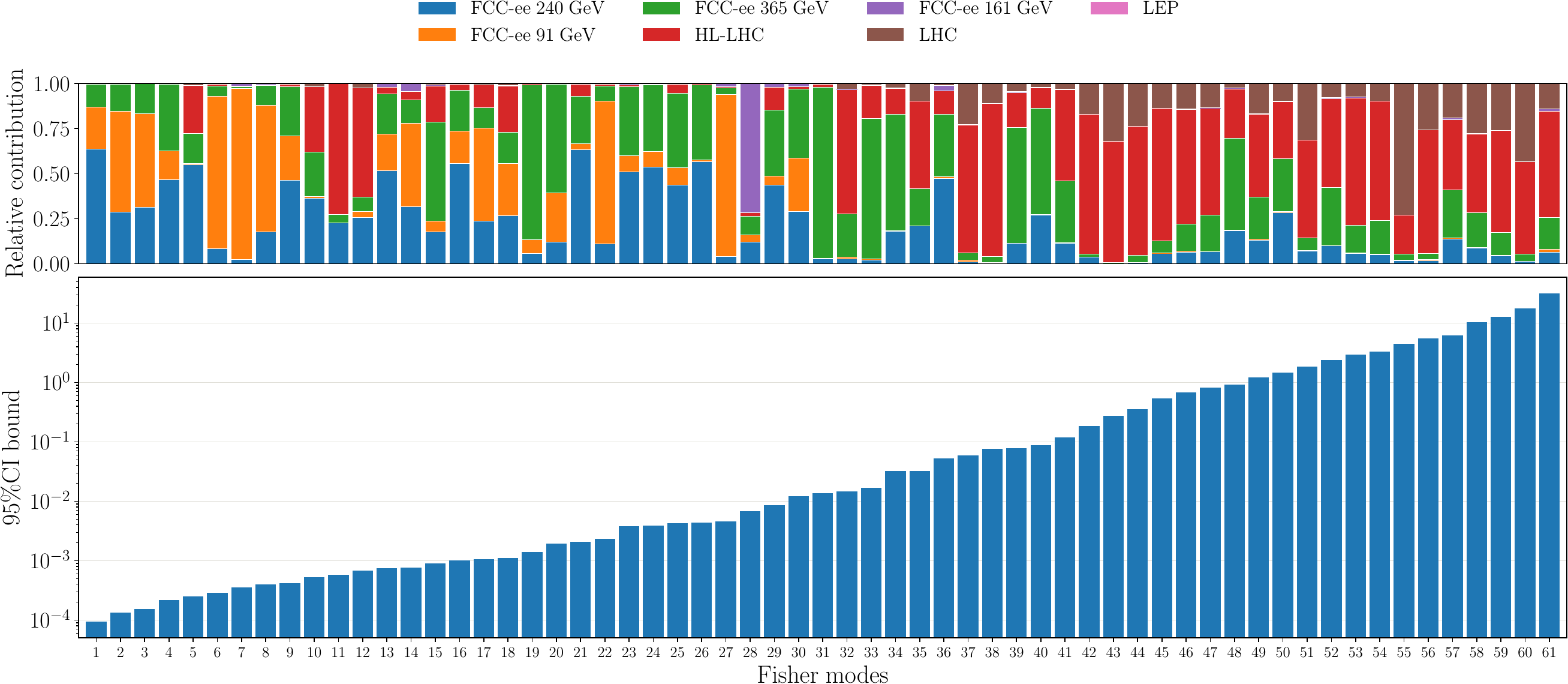}
    \caption{The relative contribution of each dataset category to the 95\% C.I bounds on the 61 principal components included in the global fit for the baseline FCC-ee scenario with aggressive theory uncertainties, while including (HL)-LHC and LEP data. The principal components are ordered from most (left) to least constrained (right). }
    \label{fig:eigenmodes_contributions}
\end{figure}

The top panel of Fig.~\ref{fig:eigenmodes_contributions} displays this breakdown in the case of the baseline FCC-ee scenario, with the corresponding 95\% C.I. bounds shown in the lower panel. Inspecting it, we observe that approximately the first 40 eigenmodes are mostly determined by FCC-ee data, while HL-LHC projections are needed in order to constrain the remaining directions~\footnote{We observe that the 55th Fisher mode is mostly constrained by the LHC rather than by the HL-LHC. The reason for this is that the main source of constraint for this direction is a mix of a $t\bar{t}\gamma$ and a dilepton $t\bar{t}$ measurements, both at 8 TeV. Given that the eigenvalue of this 55th mode is 4 orders of magnitude smaller than that of the main eigenmodes, very little impact can be expected.}. 

We provide in Table \ref{tab:rank_table} an overview of the rank, or equivalently, the number of directions which can be constrained independently, of the Fisher information matrix in various running scenarios. For example, in the FCC-ee-only fit we find that the Fisher information matrix has rank 52, and thus suffers from 9 degeneracies in our 61-dimensional space of EFT coefficients. This indicates again the important complementarity between the FCC-ee and the HL-LHC as the latter is needed to break flat directions between operators in a global EFT fit.

\begin{table}[h!]
    \renewcommand{\arraystretch}{1.3}
    \centering
    \begin{tabular}{l|c}
         \toprule
         Input dataset& Constrained directions  \\
         \midrule
         FCC-ee + LHC + HL-LHC& 61/61 \\
         FCC-ee w/o top + LHC + HL-LHC& 61/61\\
         FCC-ee only & 52/61 \\
         FCC-ee w/o top only& 38/61\\
         \bottomrule
    \end{tabular}
    \caption{The number of directions in the EFT parameter space constrained by different choices of input datasets. The LEP dataset is always included.}
    \label{tab:rank_table}
\end{table}

Fig.~\ref{fig:eigenmodes_contributions} also helps explain why the bounds obtained for the upscoped and descoped FCC-ee scenarios from Sec.~\ref{subsec:results_descoped} and ~\ref{subsec:results_upscoped} do not scale exactly with the luminosity in the marginalised setup. The HL-LHC provides a sizeable contribution to several eigenmodes, preventing the bounds along these directions from following the naive luminosity scaling.
This behaviour also holds in the original basis of EFT coefficients, since the eigenmodes dominated by HL-LHC datasets generally have non-zero components along several of the original EFT directions.

\begin{figure}[h!]
    \centering
    \includegraphics[width=\linewidth]{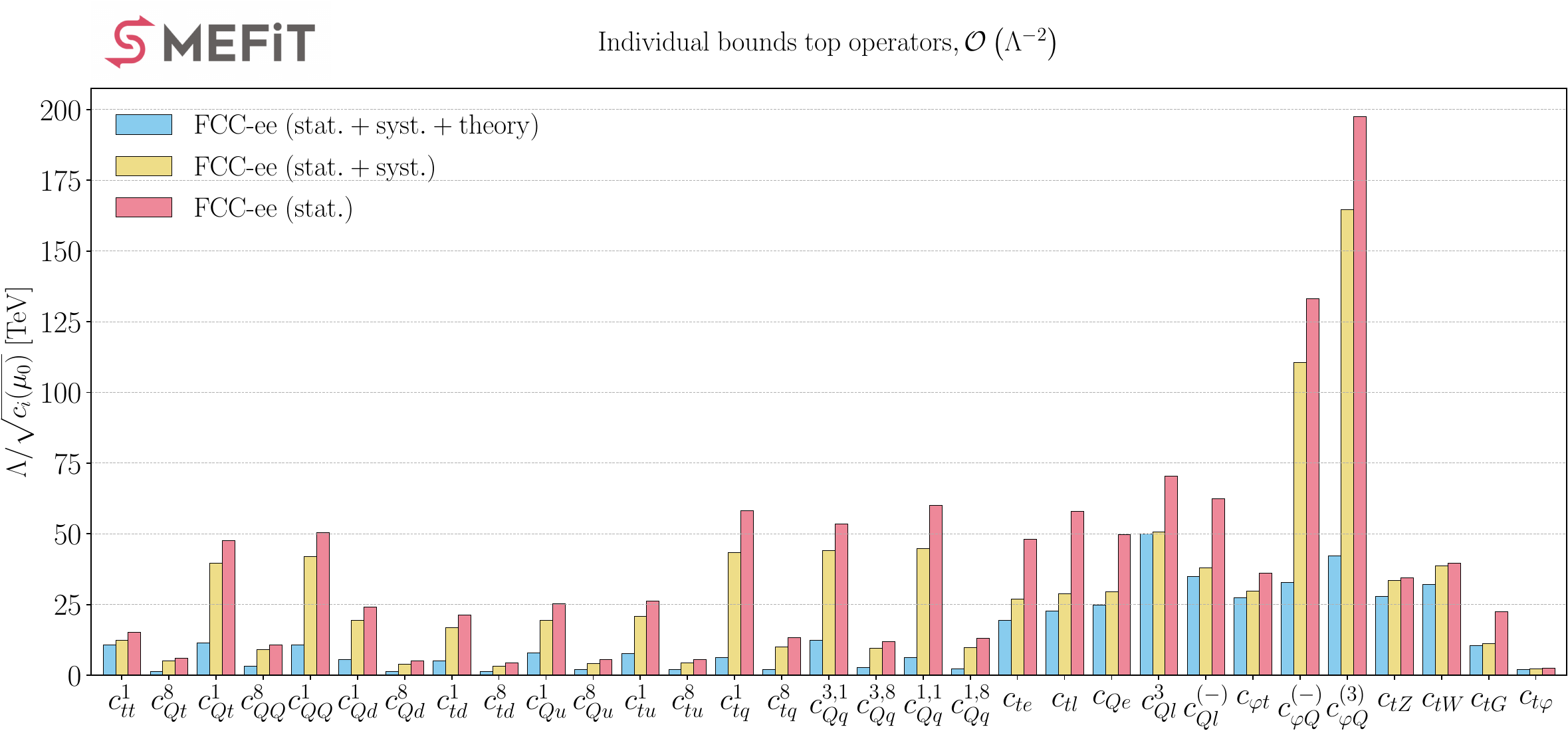}
    \includegraphics[width=\linewidth]{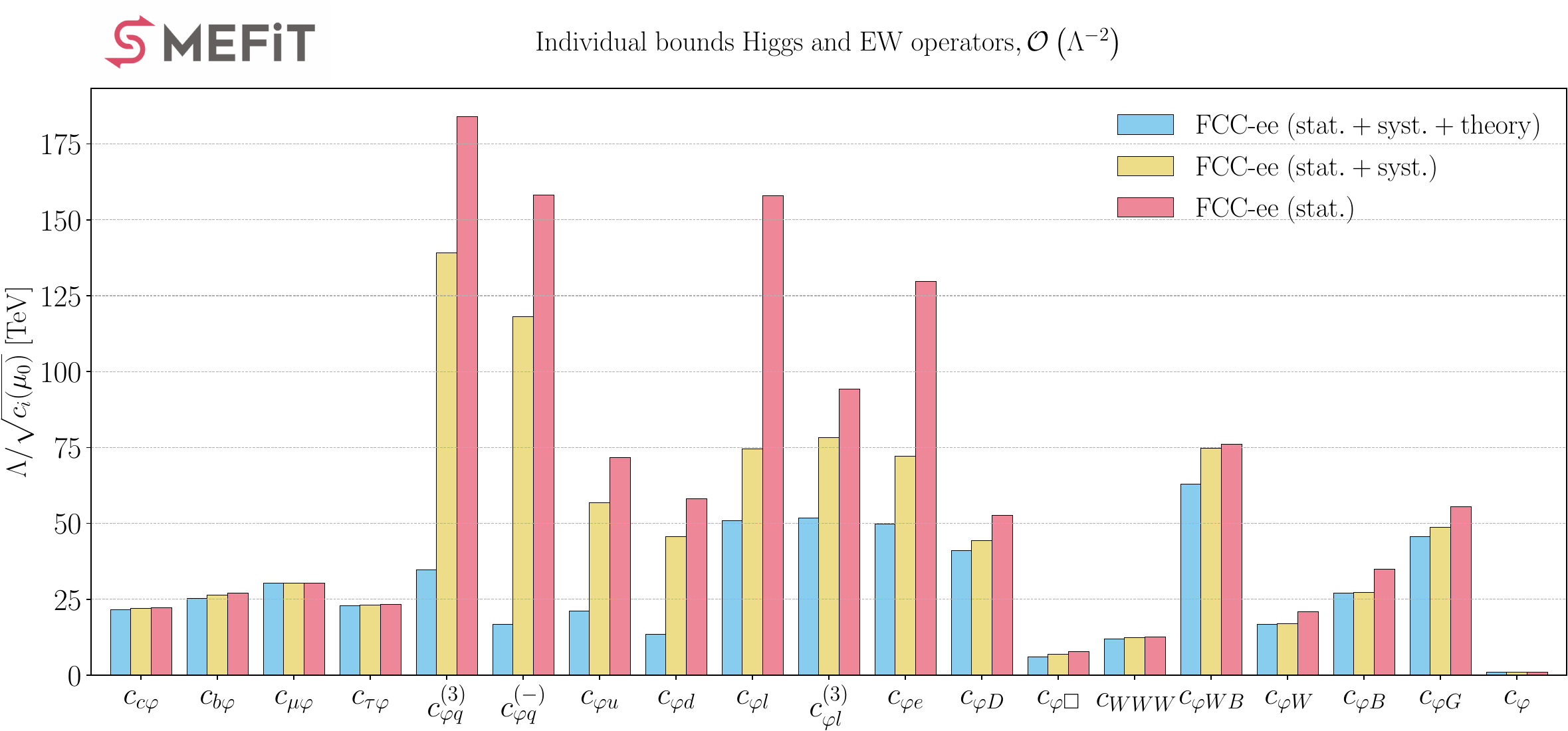}
    \caption{Comparison of the uncertainty budgets for the 95\% C.I. bounds in the case of individual fits to the baseline FCC-ee at linear order in the EFT expansion for top operators (upper panel) and Higgs and electroweak operators (lower panel). For each operator, we indicate the impact of successively removing first the aggressive theory uncertainties and then the systematic uncertainties. Parametric uncertainties are always included.}
    \label{fig:bar_unc_comp_top}
\end{figure}

\paragraph{The case of individual fits} While the global SMEFT analysis suffers from flat directions in an FCC-ee-only fit, necessitating (HL)-LHC measurements to fully constrain the parameter space, the individual fits, by definition, do not suffer from such degeneracies. To identify the dominant sources of uncertainties in individual fits, Fig.~\ref{fig:bar_unc_comp_top} displays the mass reaches of the top, Higgs, and electroweak operators in the baseline FCC-ee scenario for three different treatments of the uncertainties. Specifically, we first show the results obtained using the full covariance matrix, including statistical, systematic, parametric, and (aggressive) theory uncertainties. We then remove the theory uncertainties, retaining only the parametric, statistical and systematic experimental uncertainties, and finally remove all sources except statistical and parametric uncertainties. A first observation from Fig.~\ref{fig:bar_unc_comp_top} is that for the fermionic coefficients that enter the $Z$-pole EWPOs at tree-level, such as $c_{\varphi q}^{(3)}, c_{\varphi q}^{(-)}, c_{\varphi u}, c_{\varphi d}, c_{\varphi Q}^{(3)}, c_{\varphi Q}^{(-)} $, as well as for the four-quark operators probed through RG running to the EWPOs, the dominant limitation arises from theory uncertainties. Removing the uncertainties associated with missing higher-order corrections and theoretical inputs entering the EWPO measurements leads to a sharp improvement in the mass reaches for these coefficients, while the improvements obtained by removing additional sources of uncertainty are comparatively smaller. Leptonic coefficients such as $c_{\varphi \ell}, c_{\varphi e}, c_{t \ell}, c_{te}, c_{\varphi \ell}^{(3)} $ are affected by all sources of uncertainties and display an increase in mass reach whenever an additional source is removed.  
The other coefficients listed in Fig.~\ref{fig:bar_unc_comp_top} are primarily affected by the systematic uncertainties, as can be seen by comparing the yellow and red bars. Overall, we conclude that, in the individual fits, no single source of uncertainty dominates and that the best performance at FCC-ee will be achieved by reducing all sources of uncertainty as much as possible.

\subsection{Higgs self-coupling}
\label{subsec:higgs_self_coupling}

We now study the constraining power of the FCC-ee variants on the Higgs trilinear coupling, extending the previous SMEFiT analyses of~\cite{terHoeve:2025omu,Armadillo:2026mvp}. In the dimension-6 SMEFT, the Higgs trilinear coupling shifts from its SM value ($\delta \kappa_3 =0 $) as

\begin{equation}
    \delta\kappa_3 = -\frac{2v^4}{m_h^2} \frac{c_\varphi}{\Lambda^2} + \frac{3v^2}{\Lambda^2}\left(c_{\varphi\Box} - \frac{1}{4}c_{\varphi D}\right) \, ,
\end{equation}
where $v$ and $m_h$ denote the Higgs vacuum expectation value and its mass, respectively.

\begin{figure}[h!]
    \centering
    \includegraphics[width=0.9\linewidth]{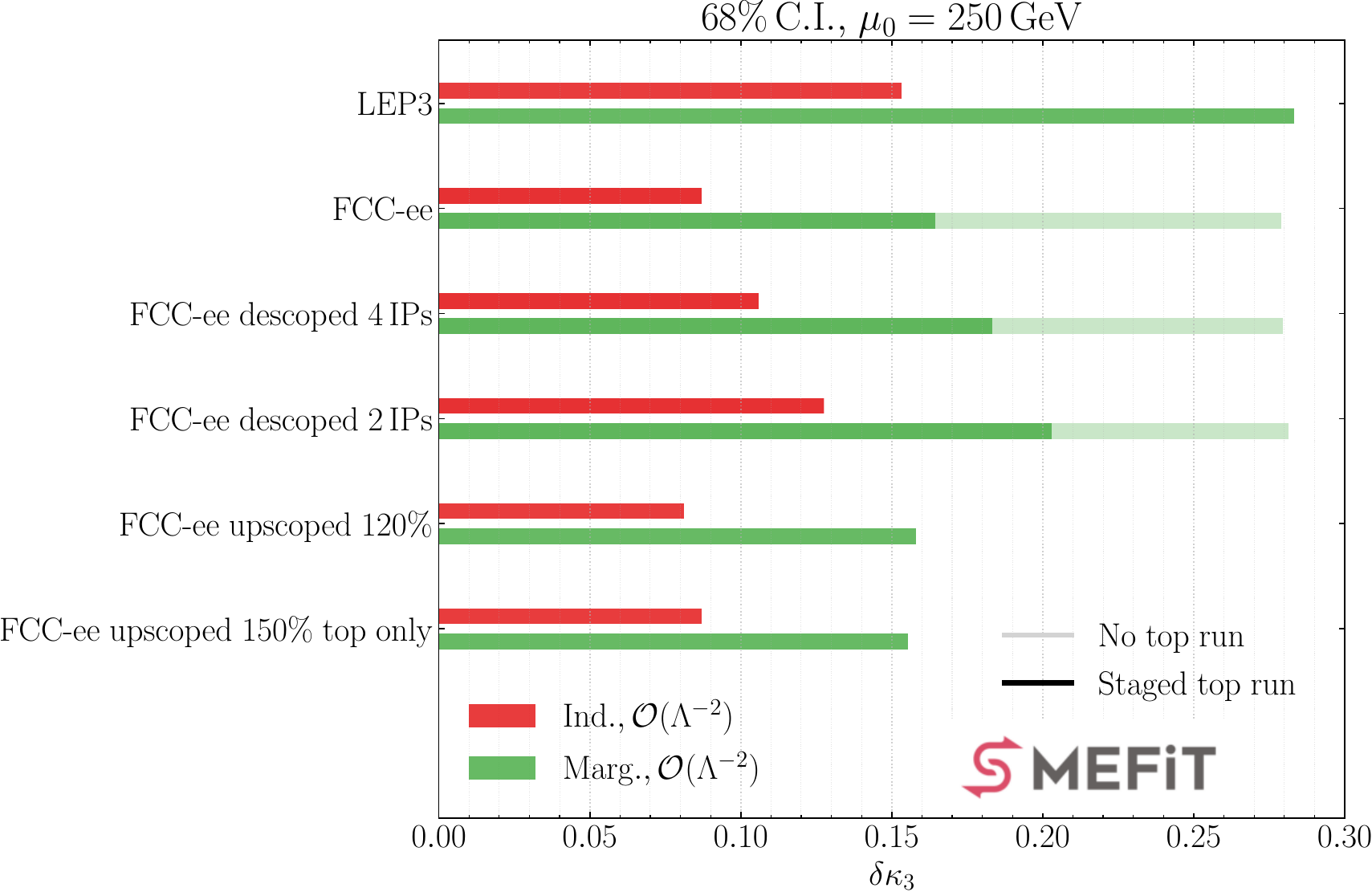}
    \caption{The 68\% C.I. on the Higgs self-coupling modifier $\delta\kappa_3$ at the linear level in the EFT expansion and at a reference scale $\mu_0=250$ GeV. We display bounds for both individual and marginalised fits, and separately show the improvement expected from the (staged) top-quark run. Aggressive theory uncertainties are adopted. }
    \label{fig:higgstrilinear}
\end{figure}

At the FCC-ee and LEP3, constraints on $\kappa_3$ can be derived from loop-induced corrections to $Zh$ production~\cite{McCullough:2013rea, Maltoni:2018ttu, Asteriadis:2024xuk, Asteriadis:2024xts, Maura:2025rcv, Allwicher:2025mvd, terHoeve:2025omu}. In Fig.~\ref{fig:higgstrilinear} we present the $68 \%$ CI bounds obtained on $\delta \kappa_3$, considering the impact of a staged top-quark run and of various integrated luminosity assignments. In this plot, the reference scale $\mu_0$ is set to the Higgs pair production threshold, a natural scale choice for the study of Higgs self-interactions. As already observed above, the run at $\sqrt{s} = 365$ GeV is crucial to outperform LEP3 in the marginalised fit: without it, both the baseline and the descoped FCC-ee scenarios lead to the same constraint on $\delta \kappa_3$ as LEP3. The top-quark run reduces the allowed deviation on $\kappa_3$ from 0.280 in the baseline, 4-IP and 2-IP descoped scenarios to 0.164, 0.183, and 0.203 respectively. 
Looking now at the two upscoped scenarios,
their performance in the marginalised bounds is slightly better than the baseline FCC-ee,
with a bound of 0.158 and 0.155 for the global upscoped 120\% and the upscoped 150\% top-only scenarios respectively. The global upscoped 120\% scenario also slightly improves the individual bound while the upscoped 150\% top-only scenario gives the same result as the baseline FCC-ee. This highlights again the predominance of the constraints coming from the $Z$-pole run in the individual fits.

\section{Conclusion}
In this work, we have studied the impact of possible descoped and upscoped FCC-ee running scenarios on the SMEFT parameter space. A descoped FCC-ee programme has been proposed by the ESPPU26 as an alternative to the full FCC-ee running programme, should the latter be financially unfeasible. Taking as our starting point the descoped FCC-ee scenario proposed by the ESPPU26, namely reduced SR power, no top-quark run, and two IPs instead of four, we have studied how these modifications to the baseline programme would affect the physics reach of FCC-ee. Motivated by the balance between SR power and operational costs, we also considered two possible upscoped FCC-ee scenarios, one in which all runs receive additional luminosity and one in which only the top-quark run does.

Several interesting findings can be highlighted from this study. First, in the absence of a top-quark run, the FCC-ee scenarios bring little improvement compared to LEP3, and they all perform similarly regardless of the integrated luminosity available at the different colliders. 
The inclusion of the top-quark run improves both the individual and marginalised constraints, although its impact is stronger in the latter case. This comes from a two-fold effect: first, the presence of additional observables helps in lifting flat directions, and second, the improved precision in the measurement of the top-quark mass leads to an indirect improvement on the parametric uncertainties. Overall, a top-quark run is essential to outperform LEP3 and fully exploit the potential of the FCC-ee. A staged top-quark run with a fraction of the luminosity of the baseline programme is sufficient to yield substantial improvements over the no-top-run scenario and should therefore be included as a minimum in any descoped FCC-ee programme.

Second, the two upscoped scenarios considered here do not significantly improve the mass reaches compared to the baseline FCC-ee programme. Combined with the observation that the descoped FCC-ee scenarios retain a physics reach close to that of the full programme once a staged top-quark run is included, this suggests that statistical experimental uncertainties may not be the dominant source of uncertainty in the SMEFT analysis. 

We thus studied the relative importance of the different sources of uncertainties included in our analysis, namely statistical and systematic experimental uncertainties, parametric uncertainties, theoretical uncertainties due to missing higher-order corrections and those associated with the theoretical inputs
required for the experimental extraction of the EWPO measurements. We found that in the case of the marginalised fit, our 61-dimensional coefficient space suffers from flat directions when considering the FCC-ee dataset alone. These flat directions are lifted by including measurements from the LHC and HL-LHC. As a consequence, the systematic uncertainties of the HL-LHC projections represent the dominant limitation in the global analysis. Following Ref.~\cite{Celada:2024mcf}, our treatment of these uncertainties is based on the projected improvement of the detectors at the HL-LHC compared to the LHC, and consists in rescaling the LHC Run 2 systematic uncertainties for selected datasets by a factor of $1/2$ for extrapolated datasets \cite{Cepeda:2019klc}, as well as some dedicated HL-LHC projections \cite{Armadillo:2026mvp}. Further developments in LHC Run 3 and HL-LHC analyses may lead to smaller uncertainties and correspondingly stronger bounds. Regardless of their precise size, our global analysis shows that the LHC should be regarded as a fundamentally complementary programme to the FCC-ee, rather than one that will be rendered obsolete by it. In the case of the individual fits, we found that no source of uncertainty clearly dominates, and the best performance of the FCC-ee will be achieved by reducing all sources of uncertainty as much as possible.

This analysis should be regarded as a first systematic assessment of the impact of a descoped FCC-ee programme. As updated projections, including dedicated datasets, and more precise predictions become available for both the HL-LHC and FCC-ee, the analysis can be progressively refined to provide a more realistic and detailed evaluation. The question of what would be an optimal set of operators to perform the fit also deserves to be explored further \cite{Mantani:2026qmg, Hirsch:2025qya}. Arguably, a physically meaningful set would be somewhere between the individual and the full global fits we have explored in this study.

\label{sec:conclusion}

\section*{Acknowledgments}
\sloppy We would like to thank the members of the SMEFiT collaboration
for insightful discussions during the course of this work, and especially Eleni Vryonidou for providing feedback on a previous version of our manuscript. We are also particularly grateful to Jorge de Blas, Christophe Grojean, Luca Mantani, Alejo Rossia and Giuseppe Ventura. E.C. is supported by the European Research Council (ERC) under the European Union’s Horizon 2020 research and innovation programme (Grant agreement No. 949451). E. H. is supported by the Swiss National Science Foundation. The work of J.t.H is supported by
the Science and Technology Facilities Council (STFC) via grant awards
ST/T000600/1 and ST/X000494/1.

\appendix
\section{Supplementary results}
\label{app:detailed_results}

\begin{figure}[h!]
    \centering
    \includegraphics[width=\linewidth]{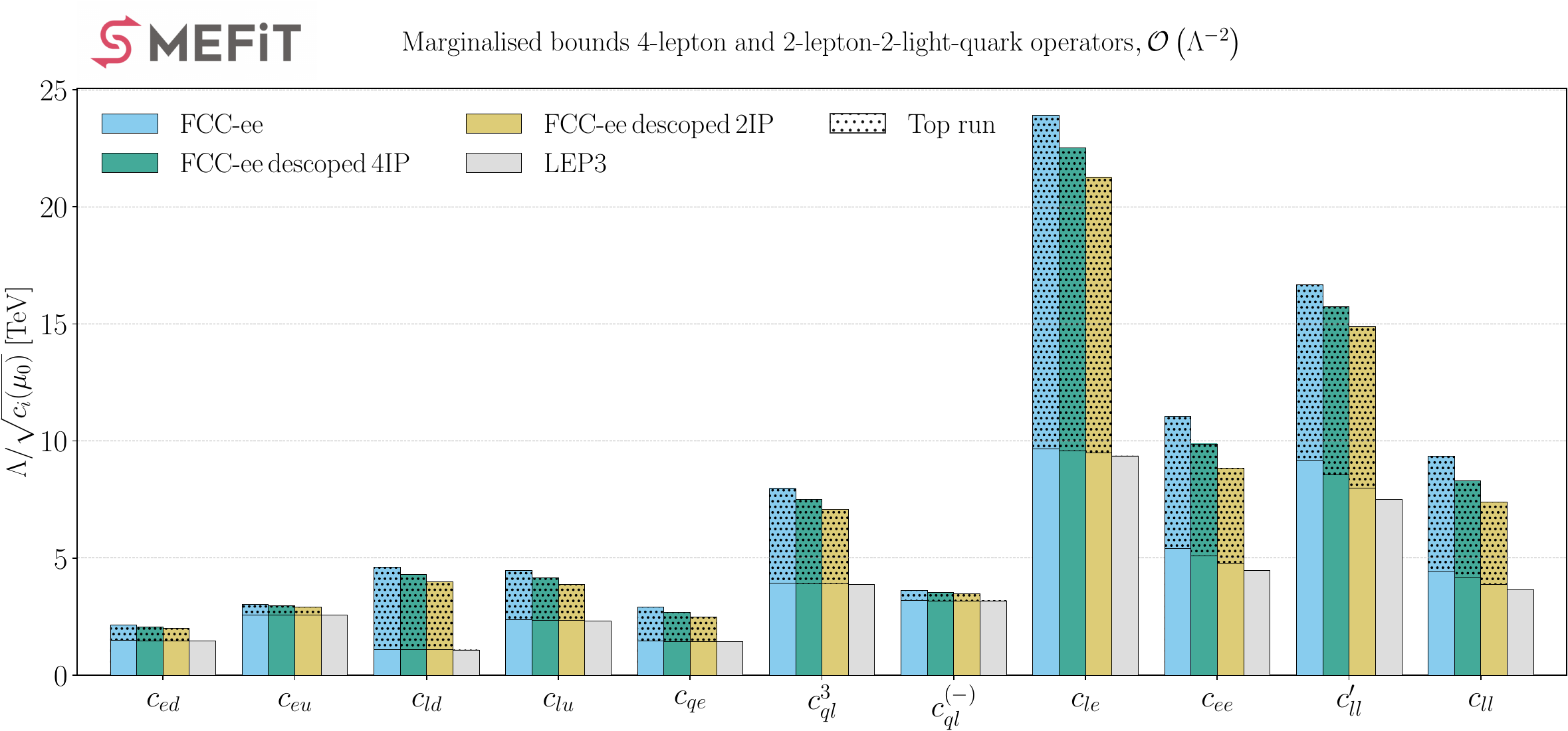}
    \includegraphics[width=\linewidth]{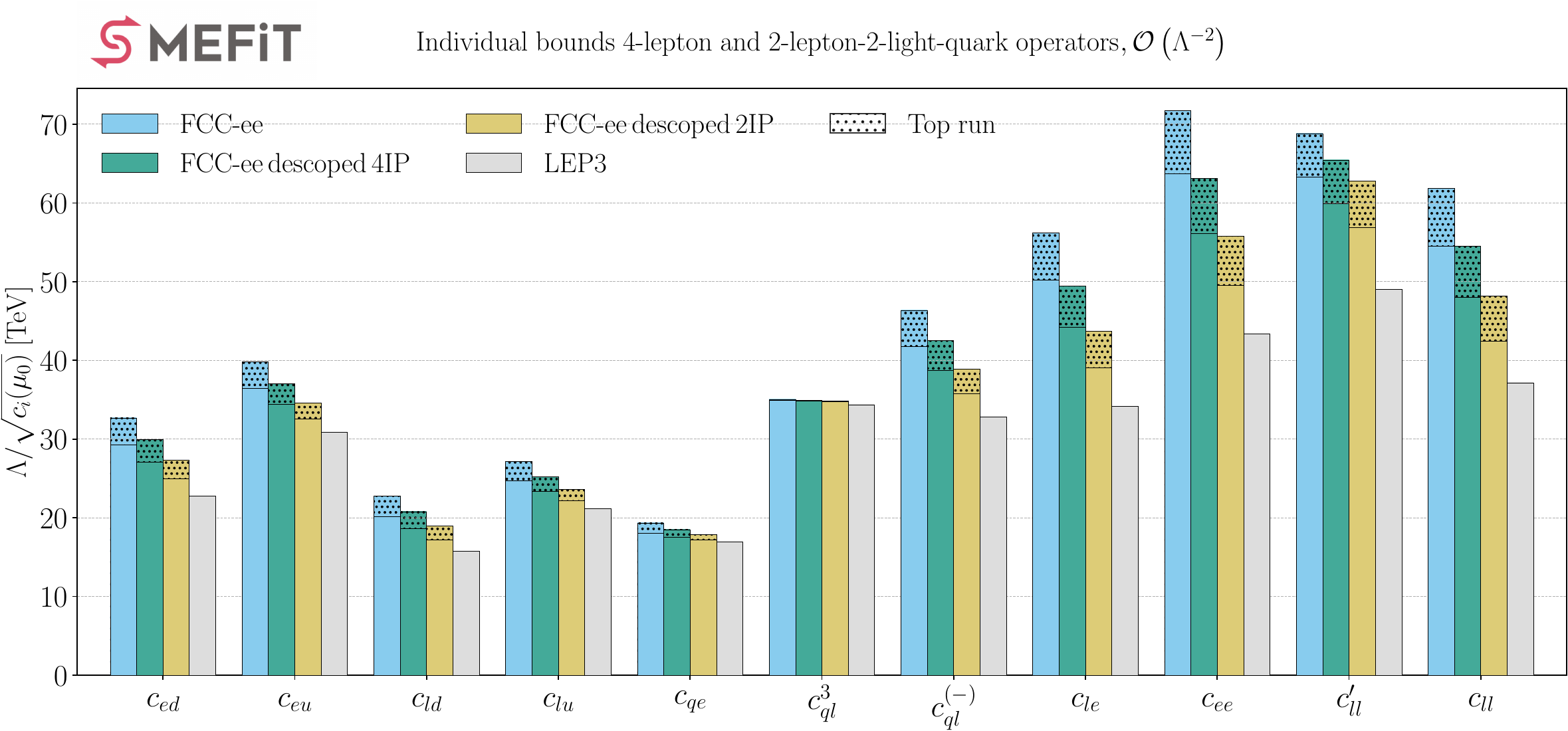}
    \caption{The 95\% C.I. on the mass reach of the 4-lepton and 2-lepton-2-light-quark operators in the marginalised (top panel) and individual (bottom panel) linear fit for the various descoping scenarios considered. The dotted bars indicate the impact of the (staged) top-quark run. Aggressive theory uncertainties are adopted.}
    \label{fig:4-lepton-2l2q}
\end{figure}

In this appendix, we display for completeness some additional results. To start with, Fig.~\ref{fig:4-lepton-2l2q} displays the mass reaches obtained for the 4-lepton and 2-lepton-2-light-quark operators that were not shown explicitly in the figures in the main text. The top panel of Fig.~\ref{fig:4-lepton-2l2q} displays the comparison between the various running scenarios for the marginalised bounds, while the bottom panel shows the corresponding individual bounds for the same set of operators. As mentioned in Sec.~\ref{subsec:results_descoped}, we find that the impact of the top-quark run is much more significant in the global fit than in the individual fit, which was also found for the other class of operators presented in the main text.

\begin{figure}[h!]
    \centering
    \includegraphics[width=\linewidth]{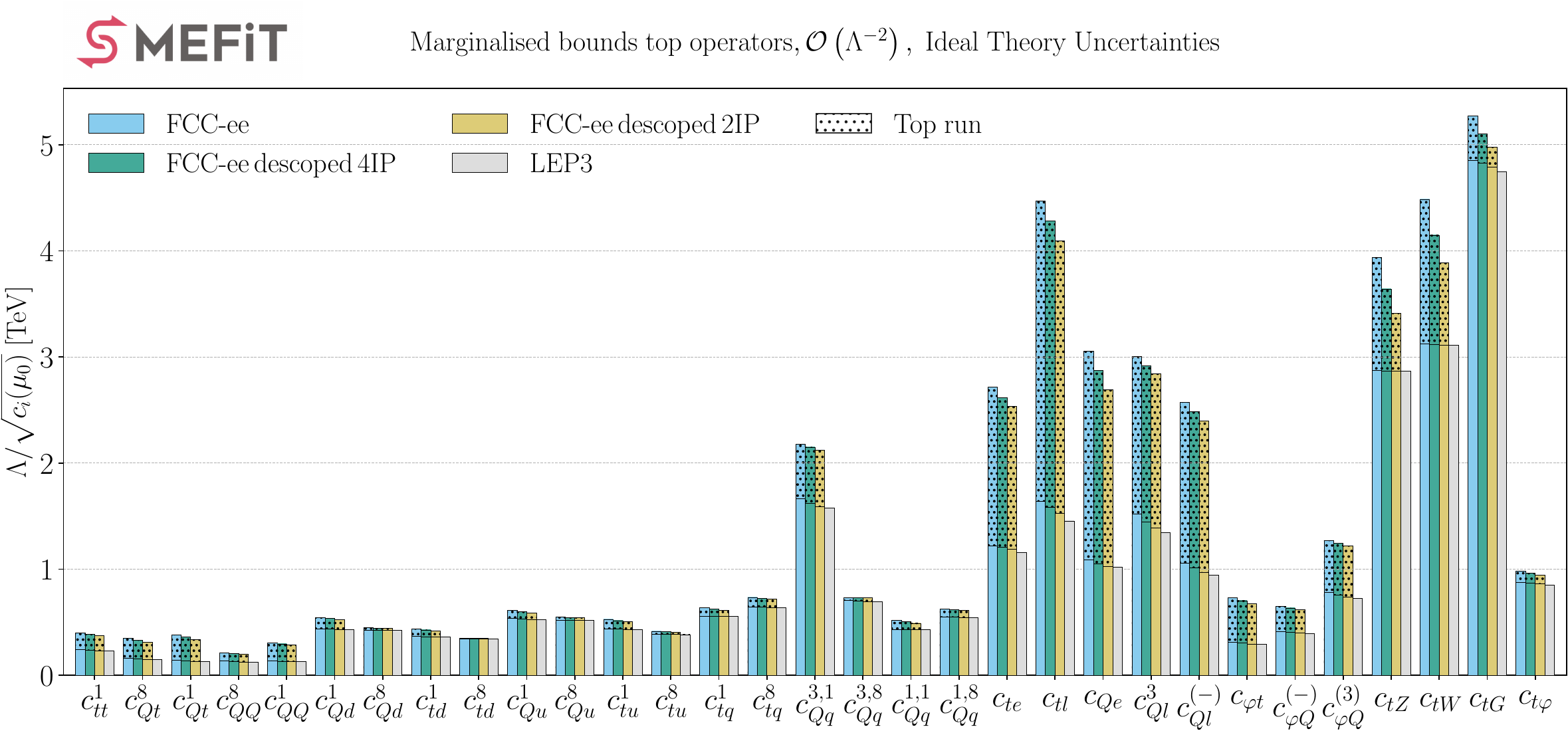}
    \includegraphics[width=\linewidth]{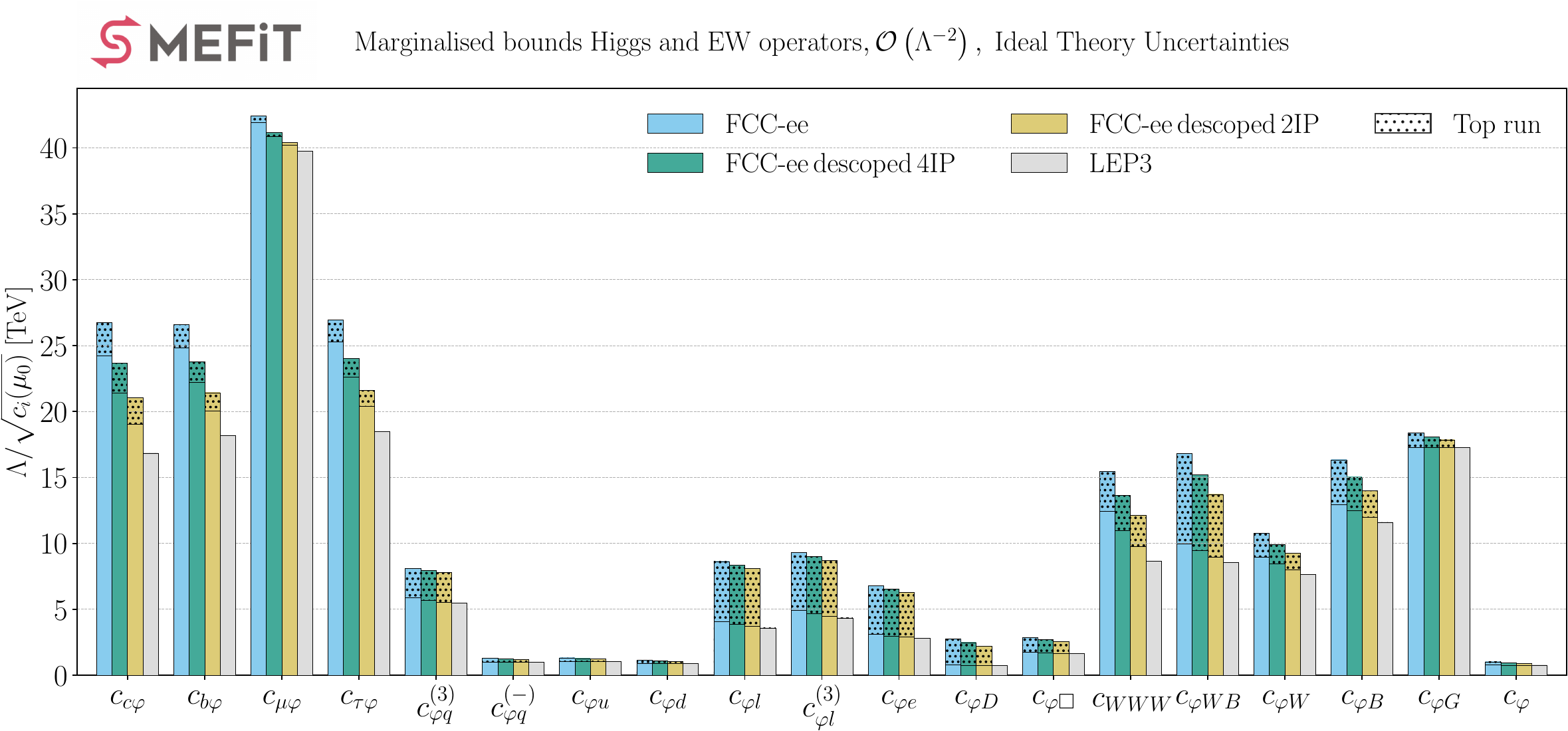}
    \caption{The 95\% C.I. on the mass reach of the top (top panel) and Higgs and electroweak operators (bottom panel) in the marginalised fit for the various descoping scenarios considered. The dotted bars indicate the impact of the (staged) top-quark run. The ideal theory scenario corresponding to the absence of theory uncertainties except for the parametric ones is adopted.}
    \label{fig:ideal_glob}
\end{figure}

Then, Fig.~\ref{fig:ideal_glob} displays the mass reaches obtained in the marginalised fit when assuming ideal theory uncertainties at the FCC-ee, LEP3, and descoped FCC-ee scenarios. As commented on in Sec.~\ref{subsec:results_descoped}, in the marginalised fit the aggressive and ideal theory scenarios yield similar results, as can be seen by comparing Figs.~\ref{fig:ideal_glob} and \ref{fig:barplot-global-aggressive} side by side.

\begin{figure}[h!]
    \centering
    \includegraphics[width=\linewidth]{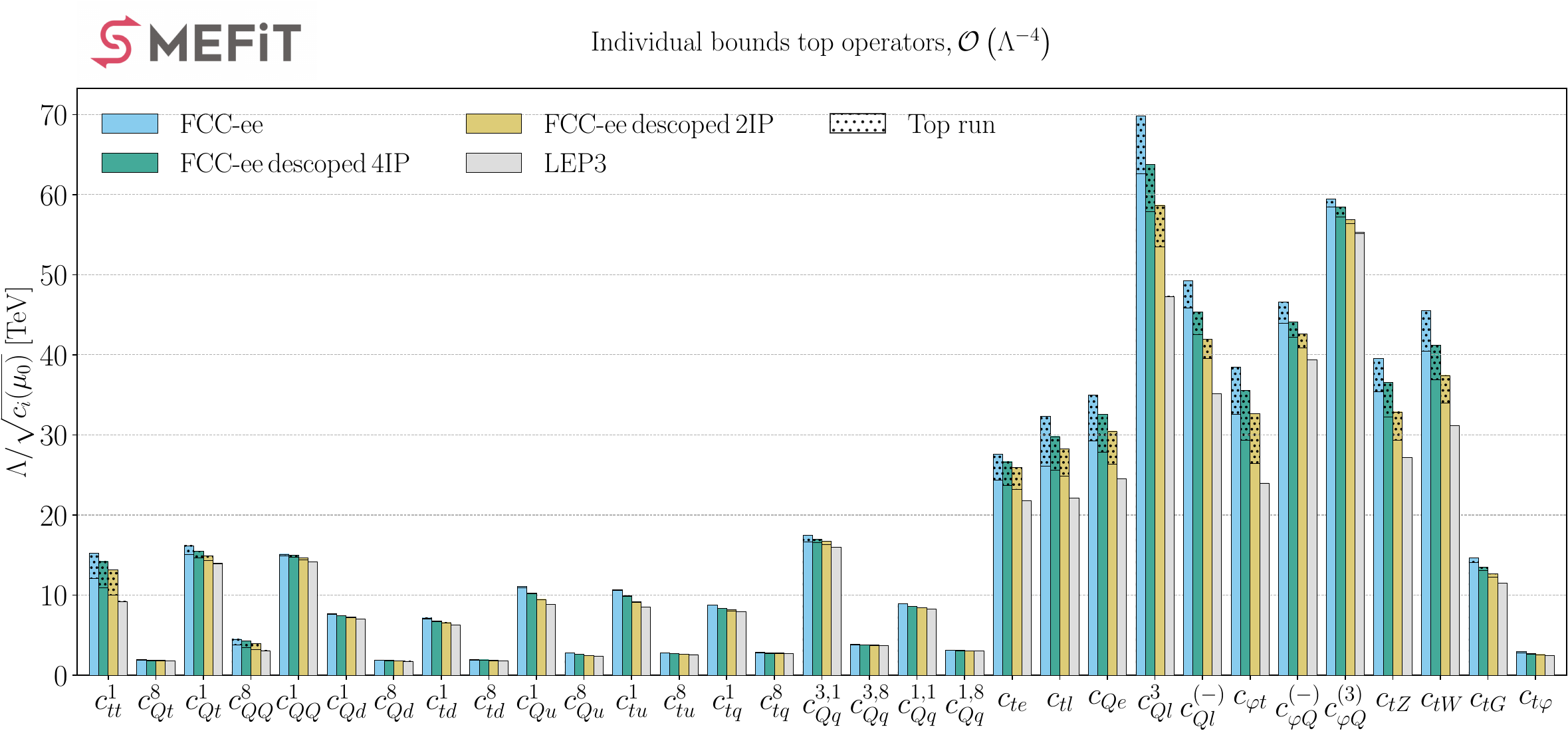}
    \includegraphics[width=\linewidth]{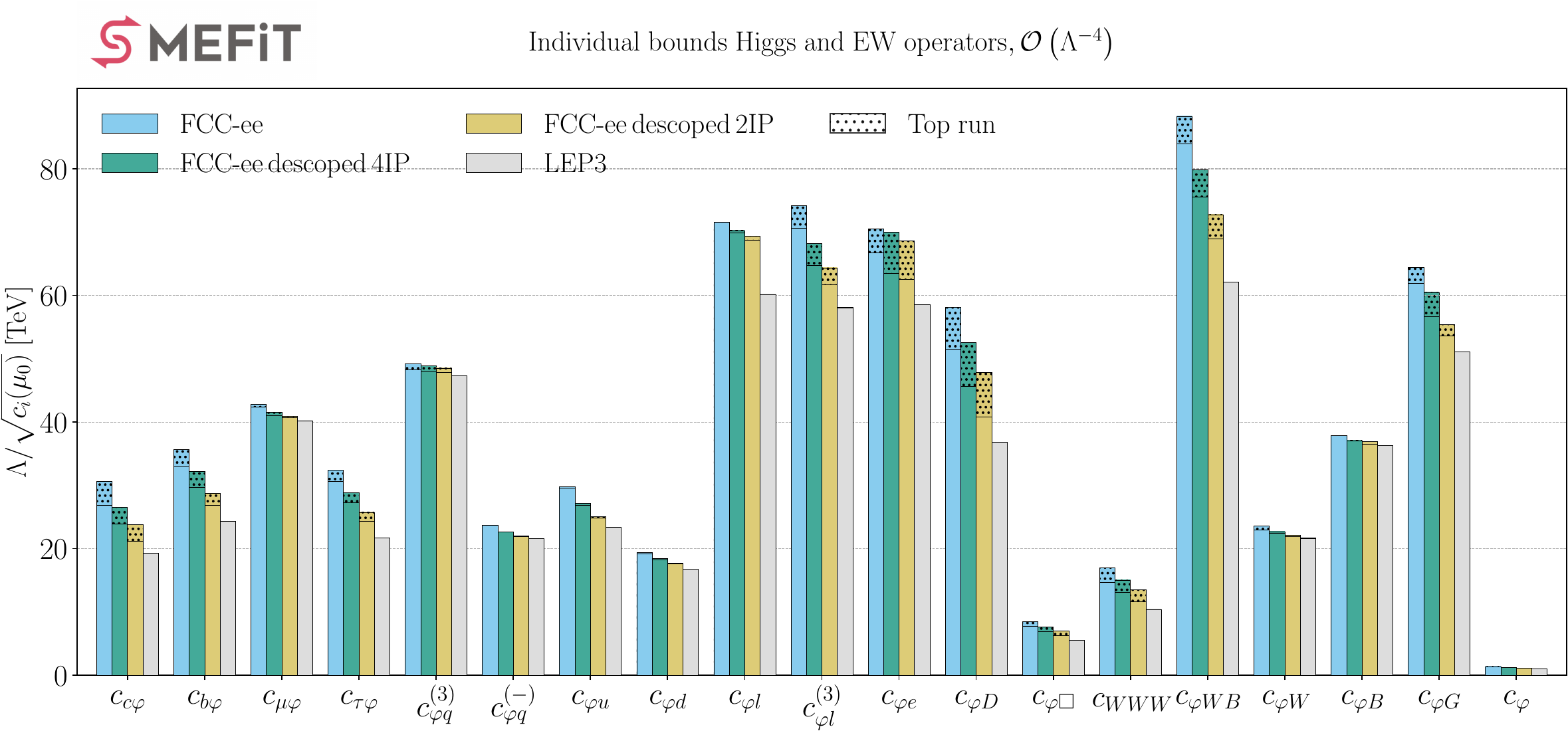}
    \caption{Same as Fig.~\ref{fig:barplot-global-aggressive-quadratic}, now for individual bounds.}
    \label{fig:barplot-ind-aggressive-quadratic}
\end{figure}

Fig.~\ref{fig:barplot-ind-aggressive-quadratic} shows the mass reaches obtained in the individual fit assuming aggressive theory uncertainties and taking into account the quadratic SMEFT corrections. We observe that the results look similar to the linear-only case in Fig.~\ref{fig:barplot-individual-aggressive}, as the individual fits are mostly driven by FCC-ee observables, particularly the $Z$-pole observables, where the linear SMEFT corrections are largely dominant.

\begin{figure}
    \centering
    \includegraphics[width=0.8\linewidth]{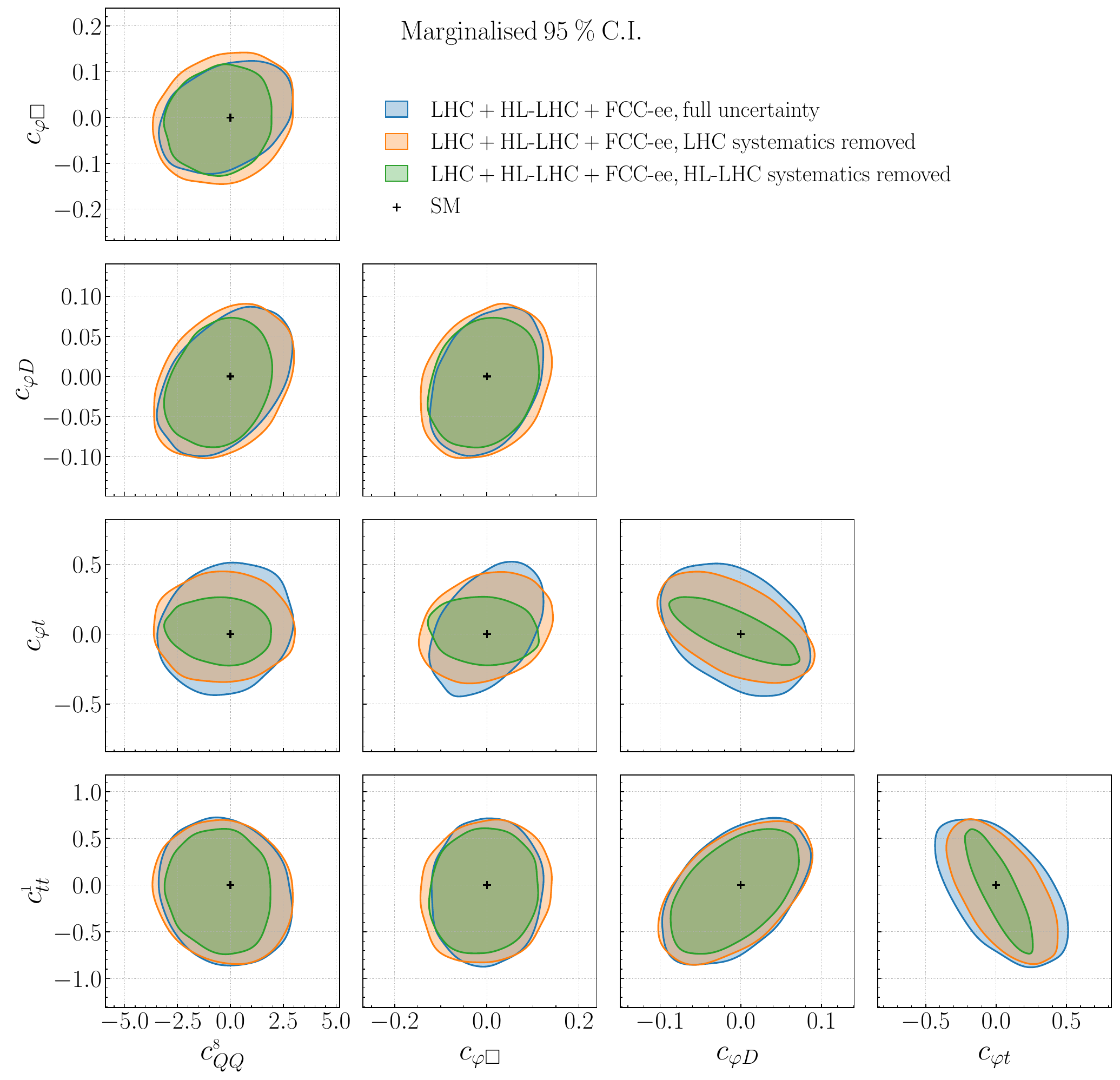}
    \caption{Same as Fig.~\ref{fig:LHC_syst_corner_plot}, now including the quadratic SMEFT corrections.}
    \label{fig:LHC_syst_corner_plot_quadratic}
\end{figure}

Fig.~\ref{fig:LHC_syst_corner_plot_quadratic} shows the impact of removing the systematic uncertainties from the LHC and HL-LHC observables when taking into account the quadratic SMEFT corrections. Although the effect is less strong than in the linear-only case from Fig.~\ref{fig:LHC_syst_corner_plot}, there is still a clear reduction of the uncertainties, suggesting that the HL-LHC remains competitive by reducing quasi-flat directions in the global fit.

\begin{figure}
    \centering
    \includegraphics[width=\linewidth]{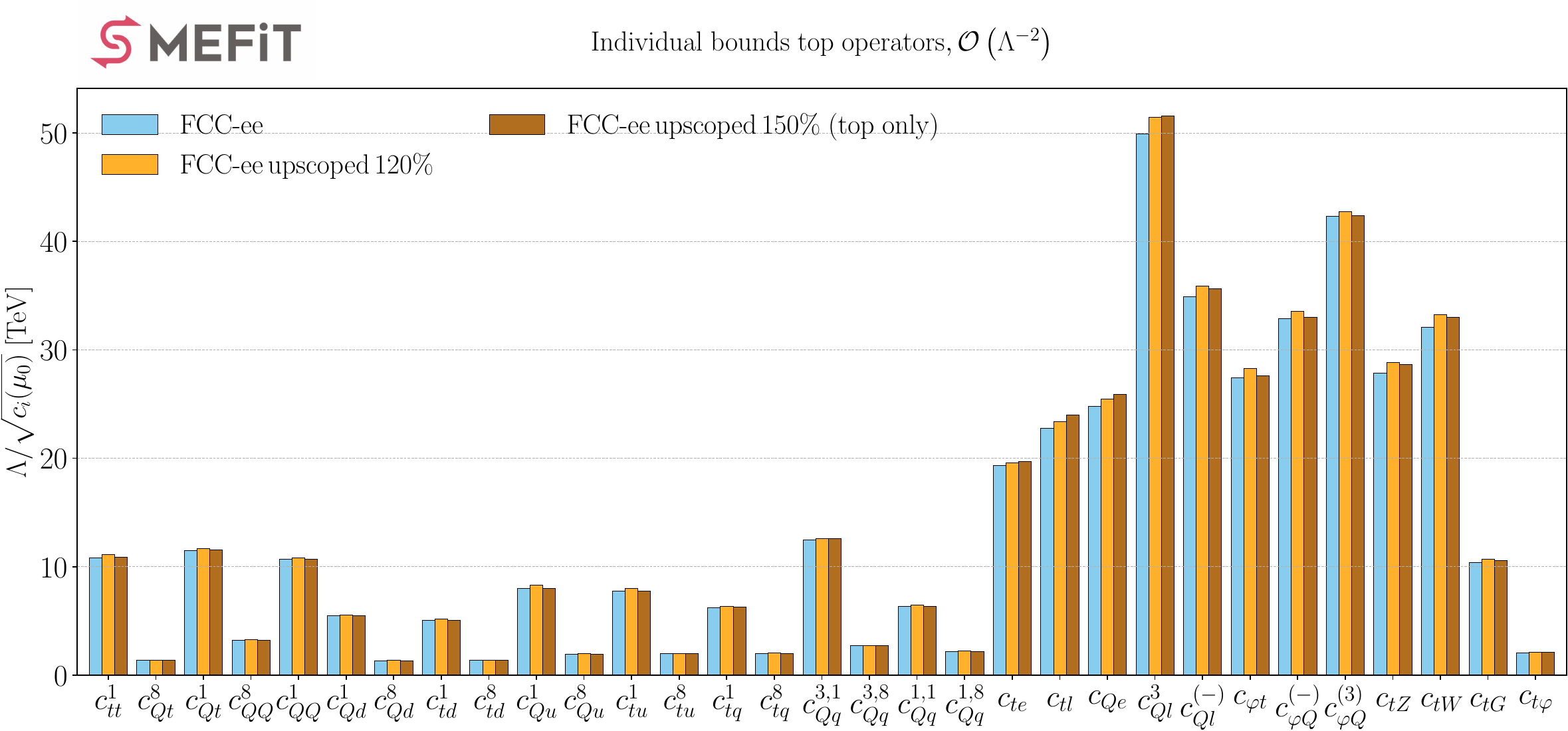}
    \includegraphics[width=\linewidth]{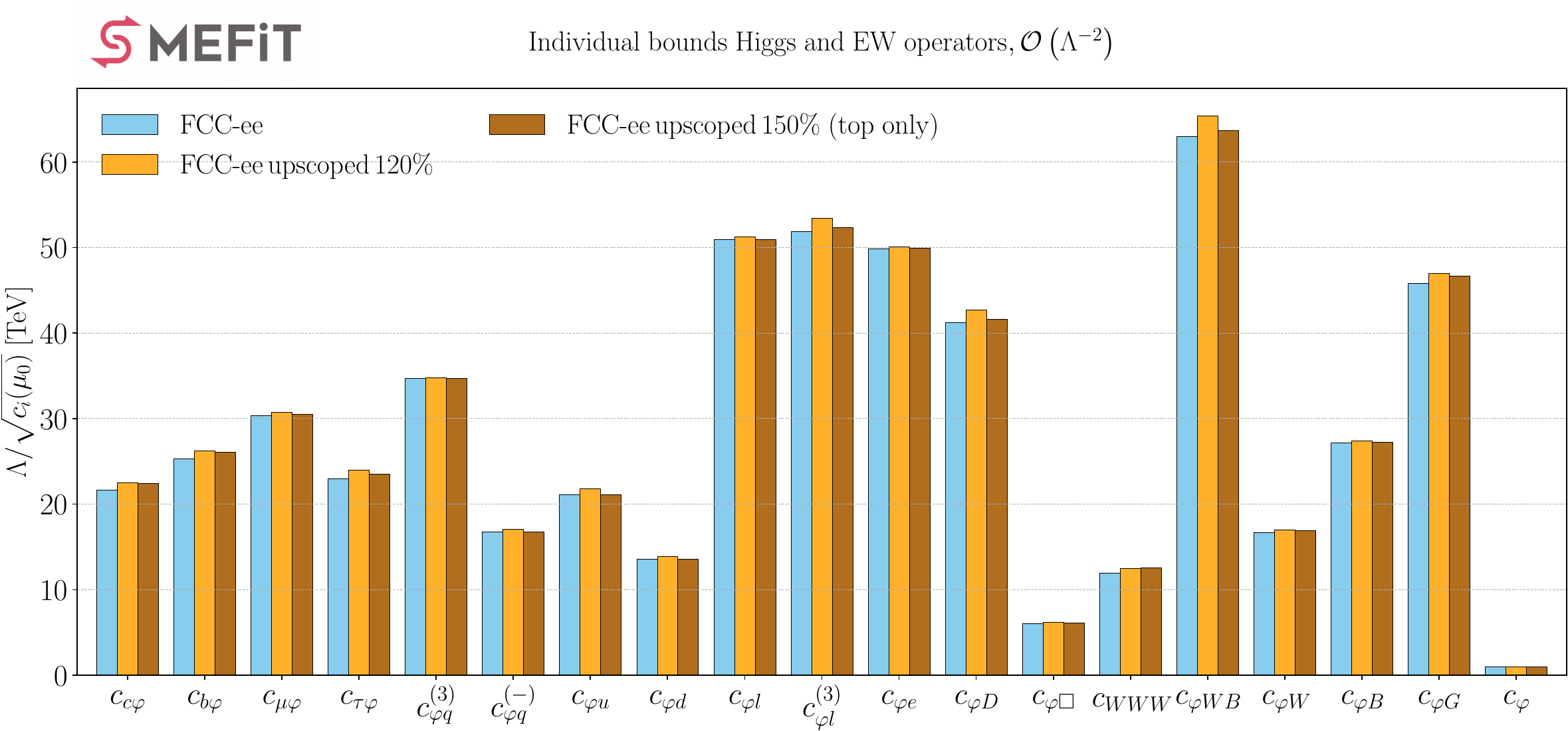}
    \caption{The 95\% C.I. on the mass reach of the top (top panel) and Higgs and electroweak operators (bottom panel) in the individual fit for the FCC-ee upscoped scenarios. Aggressive theory uncertainties are adopted.}
    \label{fig:barplot-individual-upscoped}
\end{figure}

Finally, Fig.~\ref{fig:barplot-individual-upscoped} shows the mass reaches obtained in the individual fit for the upscoped FCC-ee scenarios. Similarly to the marginalised fit case presented in Fig.~\ref{fig:barplot-global-upscoped}, the impact of the additional luminosity is limited in both upscoped scenarios.

\bibliographystyle{latex_packages/JHEP}
\bibliography{references}

\end{document}